%% file: manuscript-revision-V4.tex
\documentclass[conference,10pt, final, twocolumn, twoside, romanappendices]{IEEEtran}
\usepackage{amsmath}
\usepackage{amssymb}
\usepackage{graphicx}
\usepackage{color}
\usepackage{cite}
\usepackage{bm}
\usepackage{epsfig,psfrag}
\usepackage{tabularx}
\usepackage{acronym}
\usepackage[yyyymmdd,hhmmss]{datetime}
\usepackage{bbm}
\usepackage{mathrsfs} 
\usepackage{bardiag}
\usepackage{tabularx}
\usepackage{multirow}
\usepackage{makecell}
\usepackage{colortbl,pgfplotstable}
\usepackage{LatexInclusion/notation}
\usepackage[level=3]{LatexInclusion/messages}  % set level to 0, to hide all notes

\input{LatexInclusion/defmetric.tex}

\allowdisplaybreaks
\providecommand{\bl}{\textcolor{blue}}

\providecommand{\ist}{\hspace*{.3mm}}

\providecommand{\rmv}{\hspace*{-.3mm}}

\providecommand{\nn}{\nonumber}

\acrodef{pdf}[PDF]{probability density function}

\DeclareMathOperator*{\argmax}{arg\,max}
\DeclareMathAlphabet{\mathpzc}{OT1}{pzc}{m}{it}

\newdateformat{monthyeardate}{\monthname[\THEMONTH] \THEDAY, \THEYEAR} % specifies the format of the date

\newcommand{\paperTitle}{Navigation in Shallow Water\\ Using Passive Acoustic Ranging\\[-2mm]}
\newcommand{\paperTitleMarkboth}{}

\pgfplotsset{compat=1.14}

\begin{document}
\title{\paperTitle \vspace*{.5mm}}

\author{\IEEEauthorblockN{\large Junsu Jang and Florian Meyer \vspace{2mm}}
\IEEEauthorblockA{Scripps Institution of Oceanography and Department of Electrical and Computer Engineering} \\[-4.6mm]
\IEEEauthorblockA{University of California San Diego, La Jolla, CA}\\[-4.6mm]
Email: \{jujang, flmeyer\}@ucsd.edu \vspace{-1mm} }

\maketitle

\begin{abstract}
Passive acoustics can provide a variety of capabilities with applications in oceanographic research and maritime situational awareness. In this paper, we develop a method for the navigation of autonomous underwater vehicles (AUVs) in shallow water. Our approach relies on passively recorded signals from acoustic sources of opportunities (SOOs). By making use of the waveguide invariant, a measurement of the range to the SOO is extracted from the spectrogram of a single hydrophone. Range extraction requires knowledge of the range rate, i.e., the radial velocity between the SOO and AUV, computed from the pressure fields at different time intervals. A particle-based navigation filter fuses the range measurements with the AUV's internal velocity and heading measurements. As a result, the position error, which would otherwise increase over time, can be bounded. The ability to compute the range rate and range measurements from the pressure field measured using a single hydrophone is demonstrated on real data from the SWellEx-96 experiment. The capability of the developed navigation filter is shown based on synthetic data generated by the normal mode program KRAKEN.
\vspace{1.5mm}
\end{abstract}

\begin{IEEEkeywords}
Navigation, Bayes Filtering, Passive Acoustics, Marine Robotics, Particle Filtering.
\vspace{-4mm}
\end{IEEEkeywords}

% ----------------------------------------------
% ----------------- Sec.~I ------------------
% ----------------------------------------------

\acresetall
\section{Introduction}\label{sec:introduction}
\vspace{-1mm}

Autonomous underwater vehicles (AUVs) are widely used in scientific, commercial, and military applications due to their ability to perform complex tasks effectively. However, navigation of small and inexpensive AUVs remains a challenge. This is because underwater, the radio signals of the Global Positioning System (GPS) are unavailable. Navigation based on sensors that do not require wireless links, such as inertial sensors, relies on integrating acceleration or velocity measurements. This strategy, referred to as dead-reckoning, is subject to position errors that grow over time. As a result, the AUV resurfaces regularly to perform GPS localization \cite{GonGomCuaGarSalCab:J20}. There exist a variety of active acoustic underwater ranging techniques \cite{RogTho:J99}\cite{CheDonMilFar:J16}, but they have limited range and require deployments of multiple transponders. 

In this paper, we investigate navigation by means of a passive acoustic ranging technique. Our approach only relies on a simple acoustic sensor hosted by the AUV. Since acoustic signals emitted by a source of opportunity (SOO) are used, no additional transponders are deployed. The proposed method is applicable in shallow water with frequent ship traffic, such as coastal waterways. The SOOs are typically commercial vessels whose propeller cavitation causes the ship noise to be used as acoustic signal \cite{McKRosWigHil:J12}. In addition, this type of vessel is equipped with an automatic identification system (AIS) for broadcasting its current and predicted future position information\cite{PerChaBilKos:C09}. 

% Fig. 1 System Overview
\begin{figure}[t]
   \begin{minipage}[b]{\linewidth}
     \centering
     \includegraphics[width=8cm]{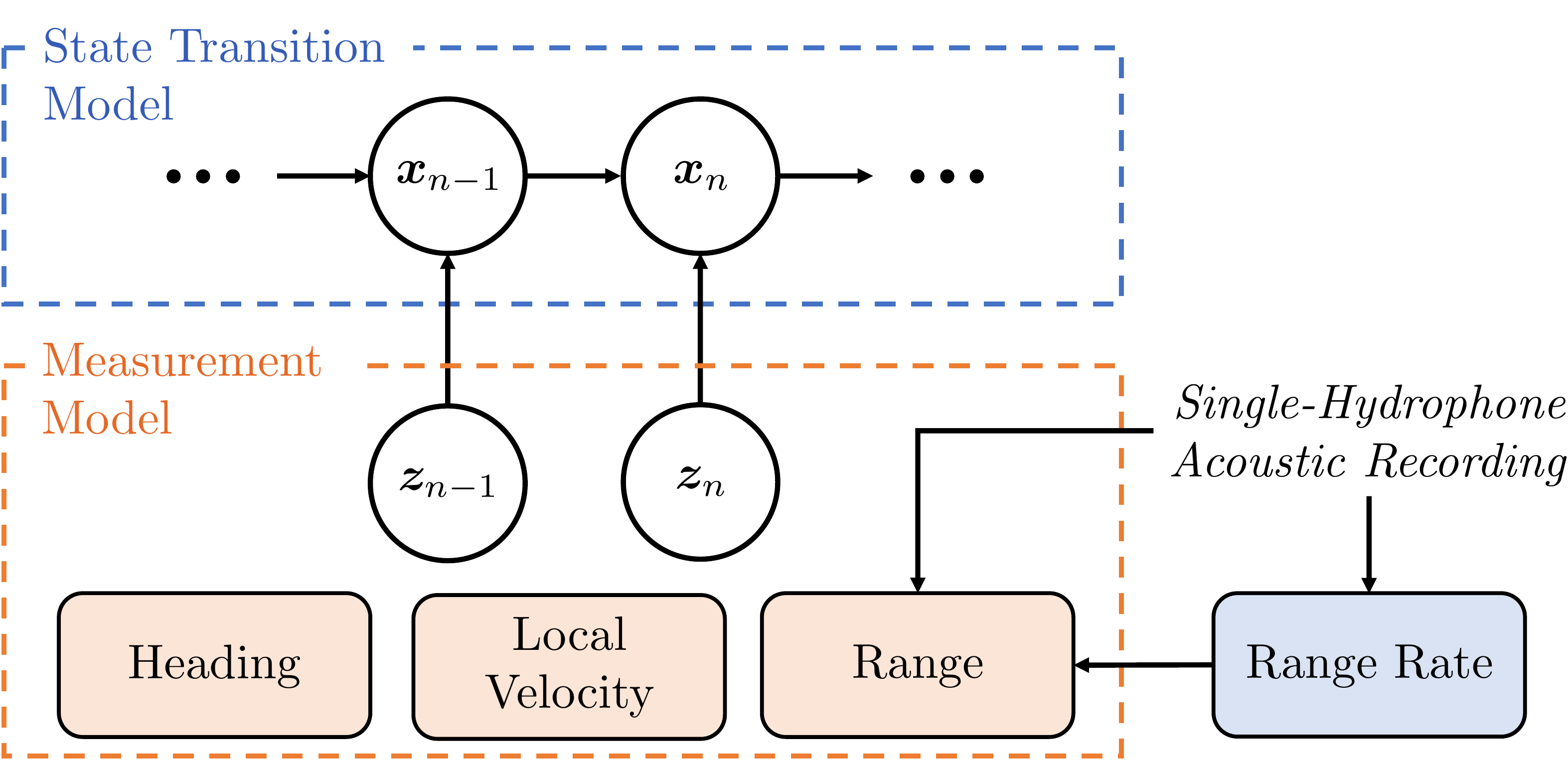}
     \vspace{-2.5mm}
     \caption{Bayesian network representing the proposed Bayes filter for AUV navigation. The state of the AUV at discrete time $n$, $\V{x}_n$, consists of the AUV's position, heading, and velocity in a global coordinate system. A measurement, $\V{z}_{n}$, consists of measurements of the range with respect to an SOO, the local velocity, and the heading. The range measurement is computed from the acoustic signal of an SOO in shallow water using the WI. The WI-based ranging relies on a range rate measurement computed in a preprocessing stage that follows the approach presented in \cite{RakKup:J12}.\vspace{-4mm}}
     \label{fig:systemOverview}
   \end{minipage}
   \vspace{-1mm}
 \end{figure}

% WI and WI Estimation methods
Our method extracts range measurements from the acoustic signal recorded by a single hydrophone. In a shallow-water waveguide, the signal emitted from a broadband moving source leads to an interference pattern in the spectrogram of a single hydrophone. In particular, minima and maxima of signal intensity are slightly shifted with respect to frequency, $f$, and range, $r$ \cite{JenKupPorSch:B11}. Loci of constant intensities form a diagonal striation pattern characterized by a scalar parameter called the waveguide invariant (WI), $\beta$\cite{Chu:J82}. The WI encapsulates the dispersive propagation characteristics of the waveguide\cite{DspKup:J99}. 
 
Parameter estimation problems in underwater acoustics that rely on striation patterns described by the WI include ranging and source localization \cite{Tho:J00,CocSch:J10,TurOrrRou:J10,ChoSonHod:J16}, geoacoustic inversion\cite{GerKinBonSteVal:J12}, and internal wave characterization\cite{KatLunOst:J16}.  In  \cite{RouSpi:J02}, the WI is treated as a probability distribution, and the impacts of environmental parameters (depth of the source and receiver, sound speed, etc...) on the WI distribution are investigated.

The WI-based ranging with respect to SOOs that exhibit tonal signals has been explored in\cite{YouSolHic:C17,YouHarHicwRogKro:J20} A single hydrophone is considered, and the application of the WI-based ranging to AUV navigation is discussed. Here, a statistical model for intensities along hypothetical striations is established, and the maximum likelihood (ML) estimation of the WI parameter and range is performed. A nonlinear least square-based approach for receiver localization is also proposed.

% Range rate
Range estimation based on the WI requires knowledge of the range rate, which can be time-varying. They are, however, assumed known in the aforementioned ranging approaches. A technique for estimating the range rate from the acoustic measurements of a single hydrophone
has been presented in \cite{RakKup:J12}. Assuming a narrowband source, this approach relies on normal mode theory\cite{JenKupPorSch:B11} to extract range rate from the spectra of consecutive snapshots of data. A combination of the range rate estimation with the WI-based ranging is also discussed. The range rate estimation has been extended to scenarios with broadband sources
in \cite{SonLiuSha:J18,ZhaSonHuiLiTan:J22}.

The fundamental question addressed in this paper is the feasibility of underwater self-localization in shallow water based on the range measurements extracted from the data of a simple acoustic receiver. In the considered scenario, tonal signals are emitted by a moving SOO. We develop a self-localization method that relies on the WI-based ranging to bound the position error of an AUV in a shallow-water environment with frequent ship traffic. Our Bayesian filter fuses the range measurements with the velocity and heading measurements provided by the internal sensors of the AUV. It offers an opportunity for an AUV, which would otherwise solely perform dead-reckoning, to recalibrate its position information without resurfacing. The WI-based ranging relies on an acoustically range-independent underwater environment, passively recorded tonal signals emitted by the SOO, and the knowledge of the potentially time-varying range rate. 

Within our approach, a range rate measurement is computed in a preparatory step \cite{RakKup:J12} by using the same tonal signal as used for the WI-based ranging. Only a single hydrophone is required to obtain a range measurement if a single SOO is present. This makes the proposed method appealing for inexpensive AUV platforms with limited acoustic sensing capabilities. We illustrate the accuracy of the range rate and range measurements using real acoustic data recorded during the SWellEx-96 experiment\cite{swellex:E96}.  We also demonstrate the capabilities of the proposed Bayesian navigation filter using acoustic data simulated in a realistic shallow water environment generated by the normal mode program KRAKEN. In particular, we show how the position error can be bounded and thus reduced compared to a reference method that relies on dead-reckoning. 

The key contributions of this paper are summarized as\vspace{.2mm} follows:
\begin{itemize}
\item We review the approaches for computing measurements of the range rate, the range, and the WI from the spectrogram recorded by a single hydrophone that is mobile with respect to an SOO.
\vspace{1.3mm}

\item We develop a Bayesian filter for navigation in shallow water that fuses information provided by the sensors for dead reckoning with the range measurements computed from the spectrogram.
\vspace{1.3mm}

\item We demonstrate passive acoustic ranging using real data from the SWellEx-96 experiment and present the capability of the developed navigation filter using synthetic data generated by the normal mode program KRAKEN.
\end{itemize}

\emph{Notation:} Random variables are displayed in serif, italic fonts, and vectors and matrices are denoted by bold lowercase and uppercase letters, respectively. 
For example, a random variable and a random vector are denoted by $x$ and $\V x$, respectively. 
Furthermore, $\|\V{x}\|$ and ${\V{x}}^{\mathrm T}$ denote the Euclidean norm and the transpose of vector $\V x$, respectively; 
$\propto$ indicates equality up to a constant factor;
$\mathpzc{p}(\V x)$ denotes the \ac{pdf} of random vector $\V x$ (this is a short notation for  $\mathpzc{p}_{\V x}(\V x)$); 
$\mathpzc{p}(\V x | \V y)$ denotes the conditional \ac{pdf} of random vector $\V x$ conditioned on random vector  $\V y$  (this is a short notation for  $\mathpzc{p}_{\V x | \V y}(\V x | \V y)$).
$\V x_{0:k}$ is short for $[\V x_0^{\mathrm T}, \dots ,\V x_k^{\mathrm T}]^{\mathrm T}$.
$\M I_n$ denotes the $n \rmv\times\rmv n$ identity matrix. 
$|\mathcal S|$ denotes the cardinality of set $\mathcal  S$.
The operator $^{*}$ denotes the complex conjugate, and $\mathpzc{j}=\sqrt{-1}$ is the imaginary unit. \vspace{0mm}

% ----------------------------------------------
% -------------- Sec.~II: WI ----------------
% ----------------------------------------------
\section{WI-Based Ranging}\label{sec:acoustics}

In this section, we will present our approach to extracting the range measurements from acoustic data by exploiting the striation pattern described by the WI. First, we discuss the preprocessing stage that will provide the range rate measurements subsequently used for the WI-based ranging. We also establish an initialization stage for the computation of the WI. In what follows, it is assumed that consecutive snapshots of data samples have been extracted from the raw acoustic signal and that the discrete Fourier transform (DFT) of length $N_{\mathrm{DFT}}$ and overlap $\alpha$ has been performed for each snapshot. The time between snapshots is  $t_{\Delta} = (1-\alpha)N_{\mathrm{DFT}} \ist T_{\Delta}$, where $T_{\Delta}$ is the sampling interval of the raw acoustic signal. Furthermore, $n = 0,1,2,\dots$ denotes the discrete-time indexes of\vspace{-.5mm} snapshots.

% Fig. range estimation steps; currently placed here for better location of this figure on the pdf
\begin{figure*}[t]
   \centering
   \begin{minipage}{0.245\textwidth}
      \centering
      \centerline{\includegraphics[width=\linewidth]{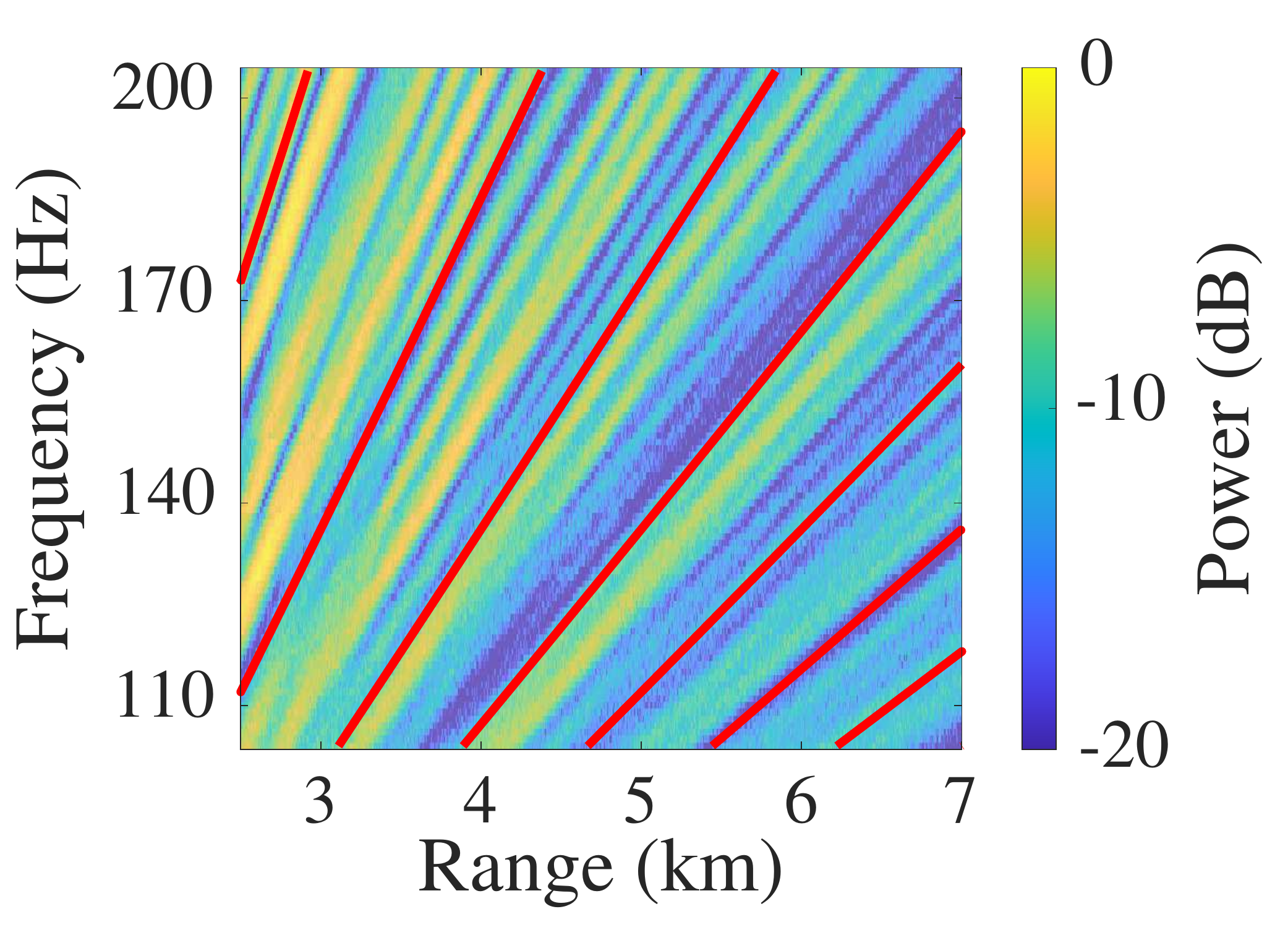}}
      \vspace{-2mm}
      \centerline{\scriptsize (a)}   
   \end{minipage}
   \begin{minipage}{0.245\textwidth}
      \centering
      \centerline{\includegraphics[width=\linewidth]{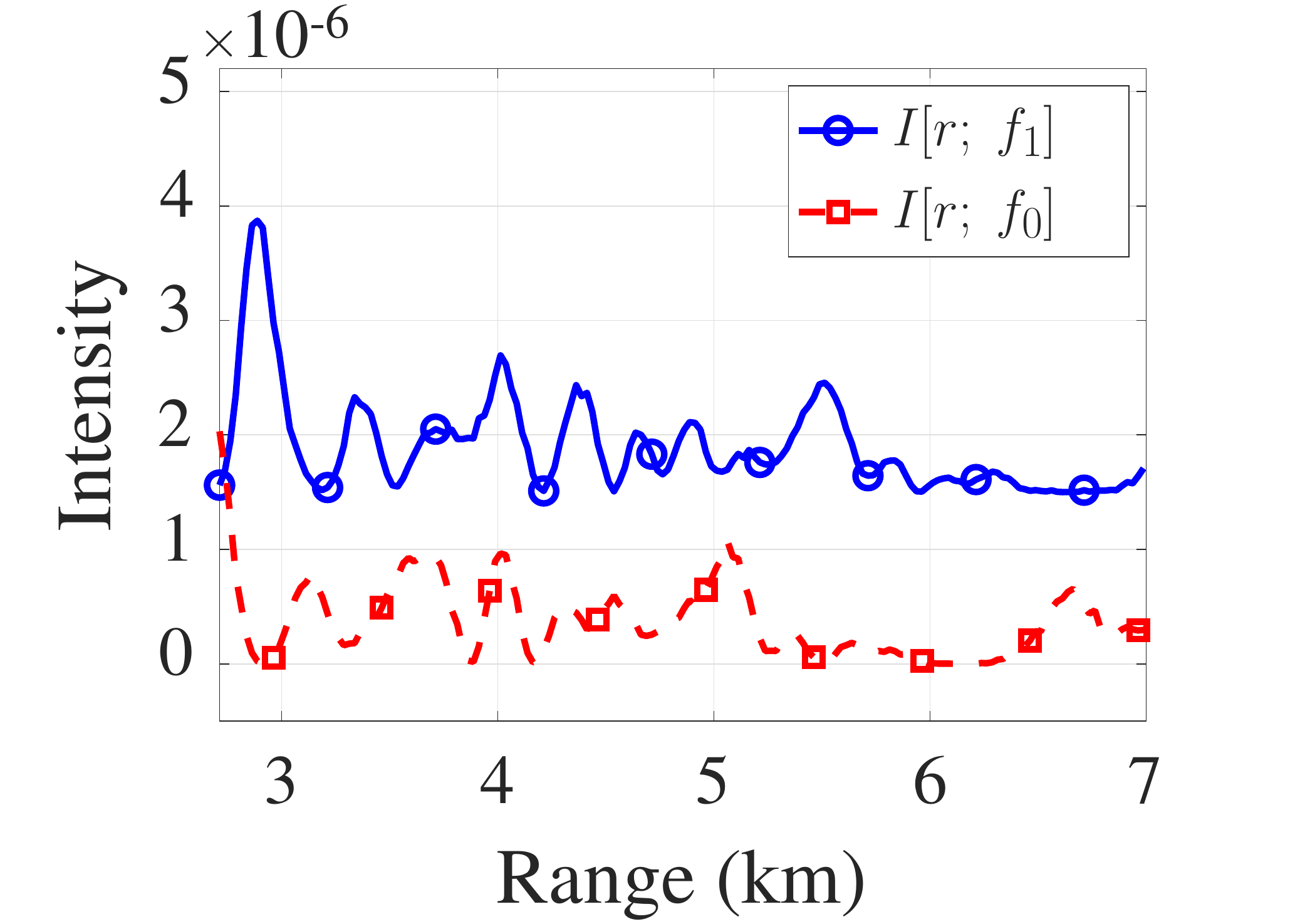}}
      \vspace{-2mm}
      \centerline{\scriptsize (b)}
   \end{minipage}
   \begin{minipage}{0.245\textwidth}
      \centering
      \centerline{\includegraphics[width=\linewidth]{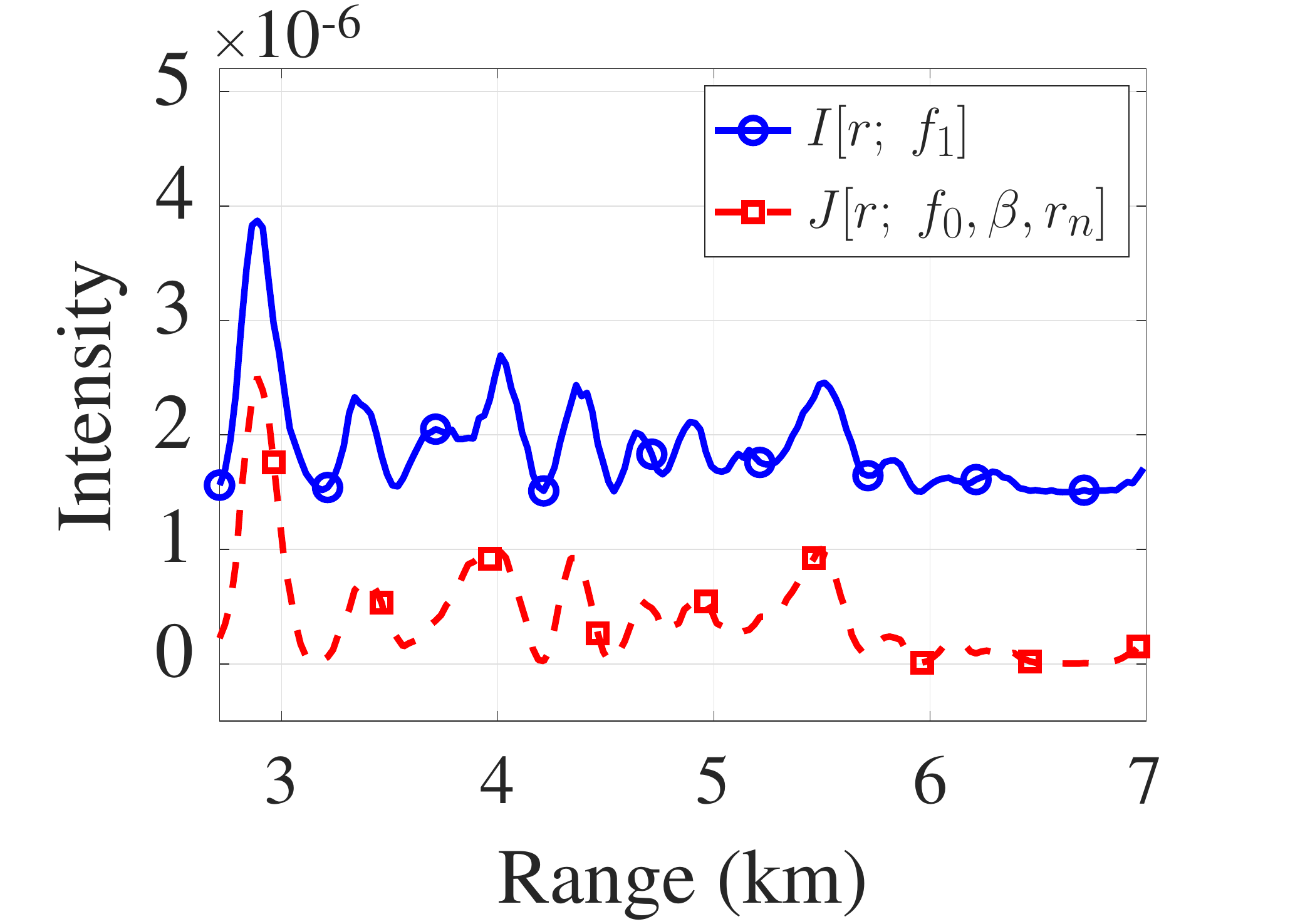}}
      \vspace{-2mm}
      \centerline{\scriptsize (c)}
   \end{minipage}
   \begin{minipage}{0.245\textwidth}
      \centering
      \centerline{\includegraphics[width=\linewidth]{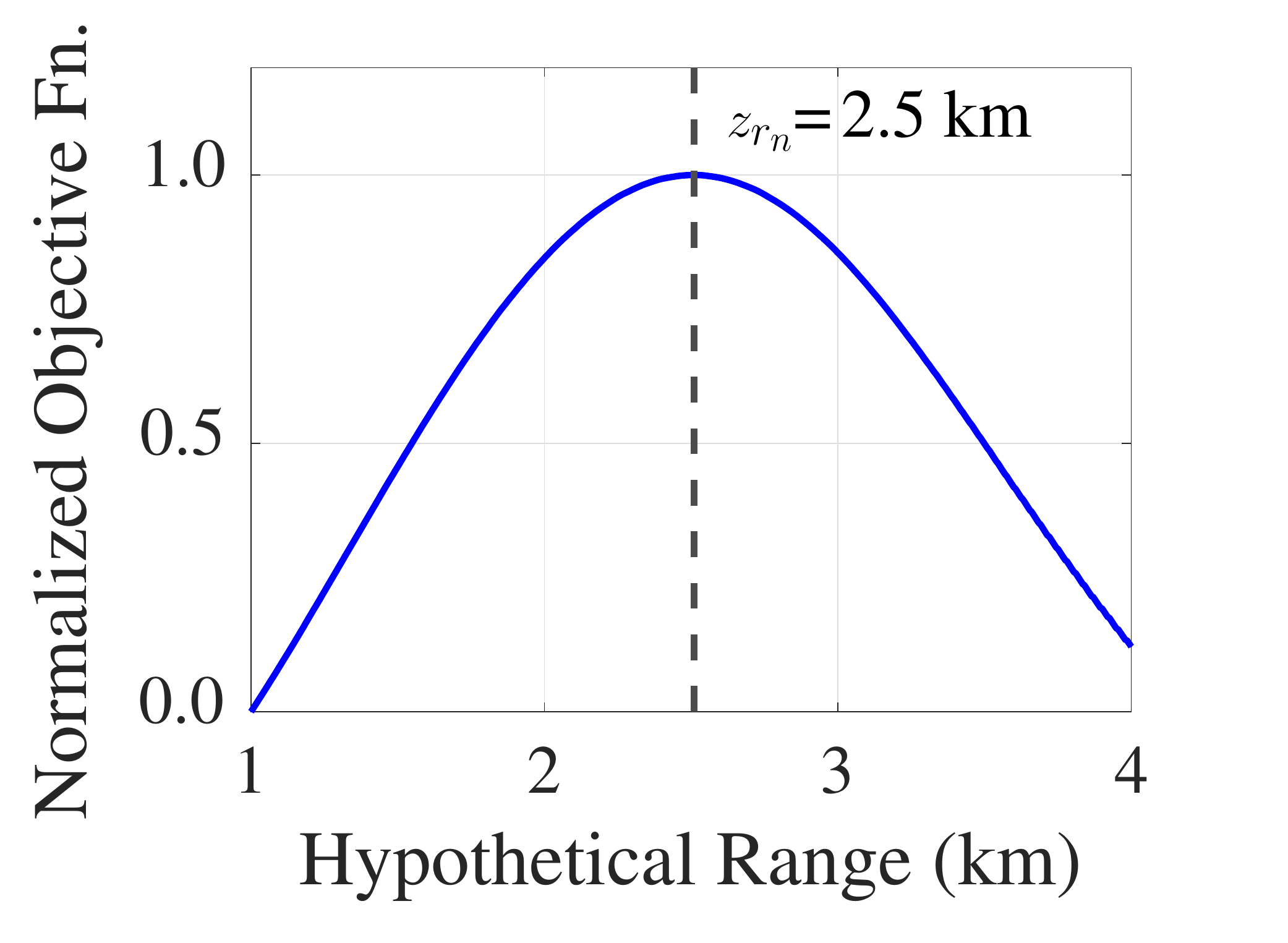}}
      \vspace{-2mm}
      \centerline{\scriptsize (d)}
   \end{minipage}
   \vspace{-3mm}
   \caption{An example of a spectrogram in the $(r,f)$ domain that represents a shallow water waveguide with $\beta = 1.07$ is presented in (a). Some expected striations are shown in red. Intensity functions at two different frequencies $f_0$ and $f_1$ with $f_0 < f_1$ are depicted as a function of range in (b) and (c), where $I[r;f_1]$ is shifted up by 1.5$\times$10\textsuperscript{-6} from its original values for visualization purposes. These intensities correspond to two different rows in the spectrogram in (a). As described by \eqref{eq:wi}, the intensity function $I[r_i;f_1]$ is a stretched version of $I[r_i;f_0]$. Based on the true $\beta$ and $r_n$, a comparison of the two intensities after stretching $I[r_i;f_0]$ by means of \eqref{eq:nonlinearTransform} is shown in (c). Here, $J[r_i;f_0,\beta,r_n]$ denotes the  stretched version of $I[r_i;f_0]$. As a result, the intensities are now aligned in range. The normalized objective function obtained by the sum of cross-correlation coefficients for pairs of demeaned intensity functions is provided in (d).\vspace{-4mm}}
   \label{fig:rEst}
   \vspace{-1.5mm}
\end{figure*}

%---------------------------------------------------------------------------%
%                       II.A. Range Rate Estimation                         %
%---------------------------------------------------------------------------%
\subsection{Computation of Range Rate Measurements}\label{sec:rrc}
At each time step $n$, we compute a measurement of the range rate, $\dot{r}_n$, following the ``difference'' method proposed in \cite{RakKup:J12}. Here, the complex DFT samples at source frequency $f$ and from consecutive snapshots of data result in the signal $p[n,f]$. This signal represents the acoustic pressure field as a function of range, $r$, and frequency, $f$. To compute a range rate measurement, we consider two segments of snapshots before and after time $n$, i.e., $p_{n,1}[k,f] = p[n \rmv-\rmv k,f]$ and $p_{n,2}[k,f] = p[n \rmv+\rmv k,f]$ with $k \rmv\in\rmv\{0,\dots,L\}$. Both segments include the sample at time $n$ and are $L+1$ snapshots long, i.e., a total of $2L+1$ snapshots are considered.

A measurement of $\dot{r}_n$ can now be extracted from the real part of the point-wise product of $p_{n,1}[k,f]$ with $p^{*}_{n,2}[k,f]$, i.e., $I_{\mathrm{c}}[k;n,f] = \mathcal{R}\big\{ p_{n,1}[k,f] \ist p^{*}_{n,2}[k,f] \big\}$. Following \cite{RakKup:J12}, the result of this product can be approximated as\vspace{.25mm} $I_{\mathrm{c}}[k;n,f] \rmv \approx \rmv \tilde{I}_{\mathrm{c}}[k;n,f]$, where $\tilde{I}_{\mathrm{c}}[k;n,f]$ is given\vspace{-.5mm} by
 \begin{equation}
   \tilde{I}_{\mathrm{c}}[k;n,f]\ist\propto\ist \cos{[ w k]}.\label{eq:crossCorr}
   \vspace{-.5mm}
\end{equation}
Here, we introduced $w \rmv\triangleq\rmv 4 \pi f \dot{r}_n t_{\Delta} / \overline{v_{\mathrm{p}}}$ where $\overline{v_{\mathrm{p}}}$ is the average horizontal phase speed \cite{JenKupPorSch:B11}. Note that $\overline{v_\mathrm{p}}$ can be further approximated by the sound speed in the water, $v$, which is assumed to be known \cite{RakKup:J12}. Eq.~\eqref{eq:crossCorr} motivates the computation of a range rate measurement, $y_{\dot{r}_n}$, from the location of largest magnitude in the frequency domain of $I_c[k;n,f]$. In particular, let $F_c[w;n,f]$ be the DFT of $I_c[k;n,f]$. A measurement of the range rate can then be obtained\vspace{-.2mm} as 
\begin{equation}
   y_{\dot{r}_n} = \argmax_{\dot{r}_n} |F_{\mathrm{c}}[4 \pi f \dot{r}_n \ist t_{\Delta} / v;n,f]|^2. \label{eq:rrMeas}
   \vspace{-1.5mm}
\end{equation}
Note that the measurement $y_{\dot{r}_n}$ represents the average range rate over the $2L+1$ snapshots. 

Unfortunately, the approach discussed above is not directly applicable to real-time navigation. This is because at time $n$, the required signal $p_{n,1}[k,f]$ consists of samples that lie in the future. The most recent range rate measurement that can be computed at time $n$ is actually the outdated $y_{\dot{r}_{n-L}}$. Thus, for real-time navigation, we use $y_{\dot{r}_{n-L}}$ to predict a real-time measurement at time $n$, denoted as $z_{\dot{r}_{n}}$. In particular, an average range acceleration measurement, $y_{\ddot{r}_{n-L}}$, is extracted from the range rate measurements $y_{\dot{r}_{n-2L}}, \dots, y_{\dot{r}_{n-L}}$. The real-time range rate measurement $z_{\dot{r}_{n}}$, is then obtained using a linear prediction\vspace{-.5mm}, i.e.,
\begin{equation}
z_{\dot{r}_{n}}= y_{\dot{r}_{n-L}}+y_{\ddot{r}_{n-L}} t_{\Delta} \ist L.
\end{equation}
Also note that the time interval $t_{\Delta}L $ has to be several minutes in length \cite{RakKup:J12}. Only in this way a range rate measurement can be computed according to \eqref{eq:rrMeas}. Thus, It is assumed that the source signal is coherent within this time interval. Furthermore, note that a phase offset is introduced when the time interval between the snapshots, $t_{\Delta}$, is not an integer multiple of $1/f$\cite{WanWanDu:J17}. This phase offset results in a shift of the peak location of $F_{\mathrm{c}}[w;n,f]$. Hence, the phase compensation approach proposed in \cite{WanWanDu:J17} is applied to obtain accurate range rate\vspace{-.5mm} measurements.

%---------------------------------------------------------------------------%
%                        II.B. range computation                            %
%---------------------------------------------------------------------------%
\subsection{Computation of Range Measurements}\label{sec:rEst}

In a shallow-water waveguide, the signal emitted from a broadband moving source leads to an interference pattern in the spectrogram of a single hydrophone. In particular, let us consider an observation interval $[t_{\mathrm{start}},t_{\mathrm{stop}}]$ and let $t_{\mathrm{start}} \leq t \leq t_{\mathrm{stop}}$ be an arbitrary time in the observation interval. We denote the ranges at times $t_{\mathrm{start}}$, $t$, and $t_{\mathrm{stop}}$ by $r_{\mathrm{start}}$, $r$, and $r_{\mathrm{stop}}$, respectively. Assuming that the source's range rate $\dot{r}$ is fixed and known during the observation interval, we can compute a spectrogram in the $(r,f)$ domain. Note that, since the range is typically unknown, the $r$-axis of this spectrogram is a function of a reference range, e.g., with $r_{\mathrm{stop}}$ being the reference range, the range at time $t$ is equal to   $r =  r_{\mathrm{stop}} - ( t_{\mathrm{stop}}-t) \dot{r}$.

This $(r,f)$ spectrogram shows an inference pattern with striations of constant intensity. Each striation can be mathematically described as\vspace{0mm}\cite{JenKupPorSch:B11}
\begin{equation}\label{eq:wi}
   \frac{f_0}{f_{1}} = \Big( \frac{r_0}{r_1} \Big)^{\beta}
   \vspace{0mm}
\end{equation}
where $r_0, r_1 \in [r_{\mathrm{start}},r_{\mathrm{stop}}] $ and $f_0,f_1$ are arbitrary source frequencies. According to \eqref{eq:wi}, the same intensity as in $(r_0,f_0)$ can be found in any other point of the spectrogram, $(r_1, f_1 )$, where $r_1 = r_0(f_1/f_0)^{1/\beta}$. The parameter $\beta$ is known as the WI. Since in typical shallow water waveguides, we have a $\beta$ close to $1$, striations often appear as diagonal lines.  An example with $\beta = 1.07$ and a known $r_{\mathrm{stop}}$ is shown in Fig.~\ref{fig:rEst}(a). The convenience of exploiting the striation pattern described by the WI for ranging is that in an approximately range-independent waveguide, all parameters that characterize the environment are summarized by the single scalar WI, $\beta$. As discussed in \cite{YouSolHic:C17}, striations are also visible when tonal signals are recorded, e.g., when a large container ship emits mechanical noise\cite{Uri:B13}. 

At time step $n$, for the WI-based ranging, we aim to compute a measurement of the range between the source and the receiver, $r_n$,  by exploiting the WI\cite{JenKupPorSch:B11}\cite{CocSch:J10}\cite{HarOdoKro:C15}. Each column in the spectrogram corresponds to discrete time step $i \rmv\in\rmv \{0,\dots, n\}$. By making use of the range rate measurements $z_{\dot{r}_{l}}$, $l \in \{ 0,\dots, n-1 \}$, computed as discussed in the previous Sec.~\ref{sec:rrc}, the following ranges can be assigned to each column of the spectrogram
\begin{equation}
   r_i =   r_n \ist\ist- \rmv\rmv\rmv\rmv \sum_{l=i}^{n-1} z_{\dot{r}_{l}}  t_{\Delta}, \hspace{5mm} i \in \{ 0,\dots, n \}.
   \label{eq:rCalc}
\end{equation}
Here, we have used $r_n$ as the reference range.

Next, we develop an objective function for the computation of $r_{n}$ by pairwise comparing nonlinearly transformed intensity functions for different frequencies. In what follows, it is assumed that the WI, $\beta$, is known. A method for the computation of a WI measurement $z_{\beta}$ will be discussed in Sec.~\ref{sec:wavInvComp}. Let $f_{0}$ and $f_{1}$ be the two source frequencies. For these two frequencies, the intensities as a function of the ranges, $r_i$, $i \in \{ 0,\dots, n \}$, are denoted as $I[r_i;f_{0}]$ and $I[r_i;f_{1}]$, respectively. Each intensity function is represented by a single row in the spectrogram.

For the WI-based ranging, we compute cross-correlation coefficients for the most recent $N$ ranges, i.e., $r_i, i \in \{ n\rmv-\rmv N\rmv+\rmv1,\dots, n \}$. Without loss of generality, let us assume that $f_{0} < f_{1}$. Based on \eqref{eq:wi} and assuming a positive $\beta$, the acoustic intensity at $f_{1}$ is a stretched function of that of $f_{0}$. An example is shown in Fig.~\ref{fig:rEst}(b). First, we perform a similar stretching for  $f_{0}$ by using \eqref{eq:wi} to convert the ranges of the last $M$ intensity samples as follows
\begin{equation}
   r'_i = r_i \Big( \frac{f_1}{f_0} \Big)^{1/\beta}\rmv\rmv\rmv\rmv, \hspace{5mm} i \in \{ n-M+1,\dots, n \}.
   \vspace{.5mm}
   \label{eq:nonlinearTransform}
\end{equation}
This results in the stretched intensity function $I'[r'_i;f_{0},\beta]$. Note that $M\rmv\geq\rmv N$ has to be chosen such that the resulting stretched intensity function is defined on the entire interval $[r_{n-N+1},r_n]$. Finally, we perform a linear interpolation to again obtain intensity for ranges $r_i$, $i \in \{ n-N+1,\dots, n \}$. This interpolation implies a cutoff of ranges outside the interval $[r_{n-N+1},r_n]$. As can easily be verified, the resulting intensity function depends on the value of the reference range $r_n$ and is thus denoted by $J[r_i;f_{0},\beta,r_n]$. An example of how an intensity function $J[r_i;f_{0},\beta,r_n]$ matches $I[r_i;f_{1}]$ after stretching and interpolation by the correct $\beta$ and $r_n$, is shown in Fig.~\ref{fig:rEst}(c). 

Given the correct $\beta$ and $r_n$, the intensity pattern, e.g., the locations of minima and maxima of $J[r_i;f_{0},\beta,r_n]$, typically matches the intensity pattern of $I[r_i;f_1]$ well. However, the absolute intensity values of the two intensity functions can be quite different, and there is a need for normalization. Therefore, the cross-correlation coefficient is used to measure similarity to develop an objective function. Let us denote the demeaned versions of $I[r_i;f_1]$ and $J[r_i;f_{0},\beta,r_n]$ as $\hat{I}[r_i;f_1]$ and $\hat{J}[r_i;f_{0},\beta,r_n]$, respectively. As an objective function for the computation of a range measurement $z_{\dot{r}_{n}}$, we now introduce the sum of the cross-correlation coefficients for all available pairs of demeaned intensity functions and the most recent $N$ samples\vspace{-.5mm}, i.e.,
\begin{align}
   g(r_n; \beta) = \sum_{(f_{0},f_{1}) \in \mathcal F} \hspace{.6mm}\frac{1}{Z}_{f_{0},f_{1}} \hspace{.5mm} \sum_{i=n-N+1}^{n} \hspace{-3mm}\hat{J}[r_i;f_{0},\beta,r_n]\ist\hat{I}[r_i;f_{1}], \nn\\[-3mm]
   \label{eq:ri} \\[-7mm]
   \nn
\end{align}
where $\mathcal F$ is the set of all pairs of source frequencies with $f_0 < f_1$ and $Z_{f_{0},f_{1}} $ is a normalization factor given\vspace{-1.2mm} by
\begin{equation}
   Z_{f_{0},f_{1}} = \Biggl[ \hspace{1mm}\sum_{i=n-N+1}^{n} \hat{J}^{2}[r_i;f_{0},\beta,r_n] \sum_{i=n-N+1}^{n} \rmv\rmv\hat{I}^2[r_i;f_{1}] \Biggr]^{\frac{1}{2}}\rmv\rmv\rmv.
\end{equation}

Finally, a range measurement is obtained by finding the location of the maximum of the objective function \eqref{eq:ri}, i.e.\vspace{.5mm},
\begin{equation}
   z_{r_n} = \argmax_{r_n} g(r_n; z_{\beta}).
   \vspace{-.5mm}
\end{equation}
Here, $z_{\beta}$ is the WI measurement to be discussed in the next Sec.~\ref{sec:wavInvComp}. An example of the objective function maximizing is shown in Fig.~\ref{fig:rEst}(d). To reduce the computational complexity, unlike the likelihood function used in \cite{YouHarHicwRogKro:J20}, the proposed objective function, \eqref{eq:ri}, does not consider noise statistics. However, as demonstrated in Sec.~\ref{sec:results}, it still leads to a competitive-ranging performance.

\subsection{Computation of the WI Measurement} 
\vspace{-1mm} 
\label{sec:wavInvComp}
A WI measurement, $z_\beta$, can be computed in an initialization stage by following a similar approach as for the computation of $z_{r_n} $. For initialization, it is assumed that the range with respect to the source, $r_n$, is known. The initialization stage is performed while the AUV remains at the surface and can record GPS measurements. In particular, for a fixed $r_n$, we obtain the WI measurement by finding the location of the maximum of $\beta$ in the objective function \eqref{eq:ri}\vspace{-.5mm}, i.e.,
\begin{equation}
   z_{\beta} = \argmax_{\beta} g(r_n; \beta).
  \vspace{-1mm}
\end{equation}
It is assumed that $\beta$ remains unchanged during the initialization stage and the next dive of the AUV. Depending on the sound speed profile, $\beta$ can change as a function of the receiver depth \cite{RouSpi:J02}. A method that can refine $\beta$ online is subject to future research.

% ----------------------------------------------
% ---------- Sec.~III: Localization ---------
% ----------------------------------------------
\section{Statistical Model and Navigation Filter}\label{sec:selfLocalization}
\vspace{-.5mm}
In what follows, we introduce the statistical model for an AUV navigation and develop the proposed navigation filter. For notational simplicity, it is assumed that the navigation filter uses the same time scale as the computation of range rate and range measurements discussed in Sec.~\ref{sec:acoustics}, i.e., each discrete time step $n$ has a duration of $t_{\Delta}$.

% Fig. 2 Track Qualitative Result 
\begin{figure}[t]
   \begin{minipage}[b]{1.0\linewidth}
     \centering
     \includegraphics[width=7cm]{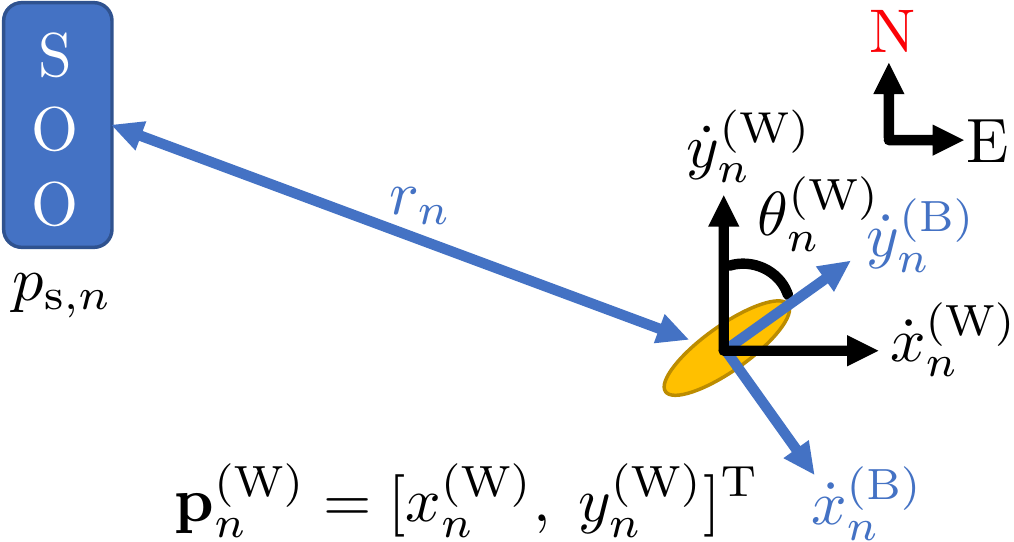}
    \caption{Relevant\vspace{.3mm} quantities for the considered AUV navigation scenario at time $n$. The state of the AUV consists of position, $\V{p}_n^{\mathrm{(W)}} = \big[x_n^{\mathrm{(W)}} \ist\ist\ist y_n^{\mathrm{(W)}}\big]^{\mathrm T}\rmv\rmv\rmv$, velocity,\vspace{.2mm} $\V{v}_n^{\mathrm{(W)}} = \big[\dot{x}_n^{\mathrm{(W)}} \ist\ist\ist \dot{y}_n^{\mathrm{(W)}}\big]^{\mathrm T}\rmv\rmv$, and heading, $\theta^{\mathrm{(W)}}_n$. The measurements required for AUV navigation are heading, $z_{\varphi_n}$, (not shown)\vspace{.3mm}, velocity in the body reference frame, $z_{\V{v}_n^{\mathrm{(B)}}}= \big[\dot{x}_n^{\mathrm{(B)}} \ist\ist\ist \dot{y}_n^{\mathrm{(B)}}\big]^{\mathrm T}\rmv\rmv\rmv$, and range between the SOO and the AUV, $z_{r_n}$. The position of the SOO, $\V{p}_{\mathrm{s},n}$, is assumed known based\vspace{.5mm} on its initial location and predicted course transmitted over the AIS.\vspace{-5mm}}
     \label{fig:states}
   \end{minipage}
 \end{figure}

The state of the\vspace{.5mm} AUV at discrete time $n$ is denoted as $\V x_n = \big[\V{p}_n^{\mathrm{(W)}} \ist\ist\ist \V{v}_n^{\mathrm{(W)}} \ist\ist\ist \rv \theta^{\mathrm{(W)}}_n\big]^{\mathrm T}\rmv\rmv\rmv$, \vspace{.2mm} where $\V{p}_n^{\mathrm{(W)}} = \big[x_n^{\mathrm{(W)}} \ist\ist\ist y_n^{\mathrm{(W)}}\big]^{\mathrm T} \rmv \in \rmv \mathbb{R}^2$ is position, $\V{v}_n^{\mathrm{(W)}} = \big[\dot{x}_n^{\mathrm{(W)}} \ist\ist\ist \dot{y}_n^{\mathrm{(W)}}\big]^{\mathrm T} \rmv \in \rmv \mathbb{R}^2$ is the\vspace{.2mm} velocity, and $\theta^{\mathrm{(W)}}_n \rmv \in \rmv  [0^{\circ},360^{\circ})$ is the heading (Fig.~\ref{fig:states}). All three quantities are defined with respect to a global (``world'') 2-D Cartesian coordinate system. In particular, the x- and y-axes are positive along the east and north directions, and the heading of 0\textdegree\hspace{1mm} points east and increases counterclockwise. The superscripts (W) and (B) indicate the ``world'' and ``body'' reference frames, respectively. The heading is relevant for the measurement model where the local velocity measurement has to be rotated from the body reference frame to the world reference frame.

The state transition between time steps follows a nearly constant velocity motion model, i.e.,
\begin{equation}
   \V{x}_n = \begin{bmatrix}
              1 & 0 & t_\Delta & 0 & 0 \\
              0 & 1 & 0 & t_\Delta & 0 \\
              0 & 0 & 1 & 0 & 0 \\
              0 & 0 & 0 & 1 & 0 \\ 
              0 & 0 & 0 & 0 & 1 
             \end{bmatrix} \V{x}_{n-1}
             + \V{u}_n
             \vspace{.5mm} \label{eq:stateTransition}
\end{equation}
where $t_{\Delta}$ is the length of each discrete time step and $\V{u}_n  \rmv \in \rmv  \mathbb{R}^{5}$ is a zero-mean multivariate Gaussian random vector with covariance matrix given\vspace{-2.3mm} by
\begin{equation}
   \M \Sigma_u =  \M{U}  \rmv\rmv\rmv\rmv\begin{bmatrix}
                 \sigma_{\V{u}_1}^2 & 0 & 0 \\
                  0 & \sigma_{\V{u}_1}^2 & 0 \\
                  0 & 0 & \sigma_{\V{u}_2}^2
                  \end{bmatrix} \rmv\rmv\M{U}^{\rmv\mathrm T}\rmv\rmv\rmv,  \quad \M U \rmv=\rmv \begin{bmatrix}
                  \frac{t_{\Delta}^2}{2} & 0 & 0 \\
                  0 & \frac{t_{\Delta}^2}{2} & 0 \\
                  t_{\Delta} & 0 & 0 \\
                  0 & t_{\Delta} & 0 \\
                  0 & 0 & t_{\Delta}
                  \end{bmatrix}\rmv\rmv\rmv.
                \nn\vspace{0mm}
\end{equation}
Here, $\sigma_{\V{u}_1}^2$ and $\sigma_{\V{u}_2}^2$ are the variance of a random linear acceleration and turn rate, respectively. The driving noise $\V{u}_n$ is assumed statistically independent across time $n$. The state-transition model in \eqref{eq:stateTransition} induces the state-transition \ac{pdf} $\mathpzc{p}(\V{x}_n|\V{x}_{n-1})$ that will be used in the prediction step of the proposed navigation filter. At the time $n \rmv = \rmv 0$, it is assumed that the prior \ac{pdf} of the AUV state, i.e., $\mathpzc{p}(\V{x}_0)$ is available.

The measurements\vspace{0mm} at time $n$ are denoted as $\V{z}_n \rmv = \rmv \big[z_{r_n}  \ist\ist\ist z_{\V{v}_n^{\mathrm{(B)}}} \ist\ist z_{\varphi_n}\big]^{\mathrm T}$\rmv\rmv\rmv\rmv\rmv, where $z_{r_n}$\vspace{-.6mm} is the range measurement that is computed as discussed in Sec.~\ref{sec:acoustics}, $z_{\V{v}_n^{\mathrm{(B)}}}= \big[\dot{x}_n^{\mathrm{(B)}} \ist\ist\ist \dot{y}_n^{\mathrm{(B)}}\big]^{\mathrm T}$ is the velocity in the local ``body'' reference frame of the vehicle, and $z_{\varphi_n}$ is a compass reading, i.e., a noisy measurement of the AUV heading. As a statistical model for the range measurements, we\vspace{0mm} use
\begin{align}
   z_{r_n} &= \|\V{p}_n^{\mathrm{(W)}} - \V{p}_{\mathrm{s},n} \| + \gamma_n. \label{eq:range}\\[-4mm]
   \nn
\end{align}
Here, $\V{p}_{\mathrm{s},n}$ is the position of the SOO in the global reference frame that is assumed known, and $\gamma_n$ is a zero-mean Gaussian measurement noise with variance $\sigma^2_{\gamma}$. The measurement noise $\gamma_n$ is assumed statistically independent across $n$. The initial position, $\V{p}_{\mathrm{s},0}$, and velocity, $\V{p}_{v,0}$, of the SOO are known, i.e., received over the automatic identification system (AIS) as the AUV starts to dive. Furthermore, it is assumed that the SOO stays on course, and thus, the position of the SOO at time $n$ can be predicted reliably, i.e., $\V{p}_{\mathrm{s},n} = \V{p}_{\mathrm{s},0} + n \ist\ist  t_{\Delta} \ist\ist \V{p}_{v,0}$. Future work will include an improved model for passive acoustic ranging that consists of a statistical characterization of the processing performed in Sec.~\ref{sec:acoustics} and takes the uncertainty of knowing the position of the SOO into account.

The velocity measurements provided by the Doppler velocity log (DVL) of the AUV are modeled\vspace{.5mm} as
\begin{align}
   z_{\V{v}_n^{\mathrm{(B)}}} &= {}_{B\rmv}R_W(\rv \theta^{\mathrm{(W)}}_n) \ist\ist \V{v}_n^{\mathrm{(W)}} + \V{\xi}_n, \label{eq:rotation}\\[-5mm] \nn
\end{align}
where $ {}_{B\rmv}R_W\big(\rv \theta^{\mathrm{(W)}}_n\big) $ is the rotation matrix from the world to the body reference frame\vspace{0mm}, i.e.,
\begin{equation}
   {}_{B\rmv}R_W\big(\rv \theta^{\mathrm{(W)}}_n\big)= \begin{bmatrix}
                    \hspace{3.4mm} \cos{\big(\theta^{\mathrm{(W)}}_n\big)} & \sin{\big(\theta^{\mathrm{(W)}}_n\big)} \\
                     -\sin{\big(\theta^{\mathrm{(W)}}_n\big)} & \cos{\big(\theta^{\mathrm{(W)}}_n\big)}
                     \end{bmatrix}\rmv\rmv
                     \vspace{0mm}\nn
\end{equation}
and $\V{\xi}_n$ is a zero-mean multivariate Gaussian measurement noise with covariance matrix $\M I_2 \sigma_{\V{\xi}}^2$. The measurement noise is assumed to be statistically independent across $n$.

Finally, the compass readings are modeled\vspace{-1.2mm} as
\begin{align}
   z_{\varphi_n} &= \theta_n + \eta_n \label{eq:velocity}   
\end{align}
where $\eta_n$ is again a zero-mean Gaussian measurement noise with variance $\sigma^2_{\eta}$. The measurement noise $\eta_n$ is also statistically independent across $n$.

The measurement models in \eqref{eq:range}, \eqref{eq:rotation}, and \eqref{eq:velocity} induce the following likelihood function at time $n$\vspace{.3mm}, i.e.,
\begin{align}
   \mathpzc{p}\big(\V{z}_n \big | \V{x}_n \big) = \mathpzc{p}\big(z_{r_n}  \big | \V{p}_n^{\mathrm{(W)}} \big) \ist \mathpzc{p}\big(z_{\V{v}_n^{\mathrm{(B)}}} \big | \V{v}_n^{\mathrm{(W)}}\rmv\rmv,  \theta_n^{\mathrm{(W)}} \big) \ist \mathpzc{p}\big( z_{\varphi_n} \big | \theta_n^{\mathrm{(W)}} \big). \nn\\[0mm]
\label{eq:likelihood}\\[-7mm]
\nn
\end{align}
This likelihood function will be used in the update step of the proposed navigation filter.

\subsection{Estimation Problem and Navigation Filter}

At time $n$, we aim to estimate the state $\V{x}_{n}$ of the AUV from all measurements $\V{z}_{1:n}$. Given the conditional \ac{pdf} of the state, $p(\V x_n|\V z_{1:n})$, the minimum mean-squared error (MMSE) estimate \cite{Kay:B93} of the state can be obtained as
\begin{equation}
   \hat{\V x}^{\mathrm{MMSE}}_{n} = \int \rmv \V x_n \ist \mathpzc{p}(\V x_{n}|\V z_{1:n}) \ist \mathrm{d}\V x_n. \label{eq:estimate}
\end{equation}
A Bayes filter\cite{AruMasGorCla:02}, which consists of prediction and update steps, is applied to compute the conditional \ac{pdf} $p(\V x_n|\V z_{1:n})$. The prediction step uses the Chapman-Kolomorogov equation, which involves the state-transition \ac{pdf}. The update step is based on Baye's rule and the likelihood function in \eqref{eq:likelihood}. Due to the nonlinearities in our measurement model, we follow the particle filtering approach \cite{AruMasGorCla:02} to compute a particle-based approximation of\vspace{-1.5mm} \eqref{eq:estimate}.

% ----------------------------------------------
% ------------ Sec.~IV: Results -------------
% ----------------------------------------------
\section{Results}\label{sec:results}
The computed range rate and range results are presented based on the simulated and real data. Then, the navigation capability is demonstrated using the simulated data. Finally, the relevant hyperparameters to the simulation are summarized in Table.~\ref{tab:hyperparams}.
\begin{table}
   \renewcommand{\arraystretch}{1.5} 
   \begin{center}
   \resizebox{\linewidth}{!}{
   \begin{tabular}{  p{6.5cm} p{1.5cm} }
   \hline
   \bf Hyperparameters  & \bf Value \\
   \hline
   \rowcolor{blue!10!white}Prior Position STD (m)                                                  & $500$ \\
   Driving Noise STD, $\sigma_{\V{u}_1}$ (m$\,$s\textsuperscript{-1})                             & $0.02$ \\
   \rowcolor{blue!10!white}Turning Rate Noise STD, $\sigma_{\V{u}_2}$ (\textdegree\hspace{.5mm})  & $0.50$ \\
   Measurement Noise (Heading) STD, $\sigma_{\eta}$ (\textdegree\hspace{.5mm})                & $0.50$ \\
   \rowcolor{blue!10!white}Measurement Noise (Range) STD, $\sigma_{\gamma}$ (m)                    & $150$ \\
   Measurement Noise (Velocity) STD, $\sigma_{\V{\xi}}$ (m$\,$s\textsuperscript{-1})                & $0.02$ \\
   \rowcolor{blue!10!white}Average Acoustic SNR for navigation(dB)                                  & $12$ \\
   Number of Particles                                                                             & $500,\rmv000$ \\      
   \hline
\end{tabular}}
\end{center}
\caption{Hyperparameters used to simulate the proposed AUV navigation method.}
\label{tab:hyperparams}
\vspace{-7.5mm}
\end{table}
For the investigation using the real data, the acoustic recordings on the first element of the Horizontal Line Array South (HLAS) are processed from the S5 event of the SWellEx-96 experiment (Fig.~\ref{fig:swellexResult}(a)). The acoustic data from 23:52 to 00:29 in Greenwich Mean Time (GMT) are used. The sampling frequency is 3276.8 Hz, and the spectrogram was generated using a window length of 3276 samples (no overlap, Hanning\vspace{-1mm} window).

\subsection{Simulation Environment and Data}\label{sec:simScenario}
\vspace{-.5mm}
As for the simulation, a scenario where a moving AUV, equipped with a single hydrophone, a DVL, and a compass, records a sound from a moving SOO in a shallow-water environment is considered. Initially, the SOO is at $(2000, 0)$ moving north at 10 knots (${\sim}5.14$ m$\,$s\textsuperscript{-1}). The AUV starts at the origin, moving northeast at 3knots (${\sim}1.54$ m$\,$s\textsuperscript{-1}), i.e., $\V x_0 = [0,0,1.1,1.1,45]^{\mathrm T}$, for 20 minutes. Its driving noise, $\sigma_{\V{u}_1}$, is 0.02 m$\,$s\textsuperscript{-1}, which accounts for random motion due to the control and the current in the ocean, and the turning rate noise, $\sigma_{\V{u}_2}$, is 0.5\textdegree\hspace{1mm}. 

\begin{figure}[t]
   \begin{minipage}[b]{1.0\linewidth}
     \centering
     \includegraphics[width=7cm]{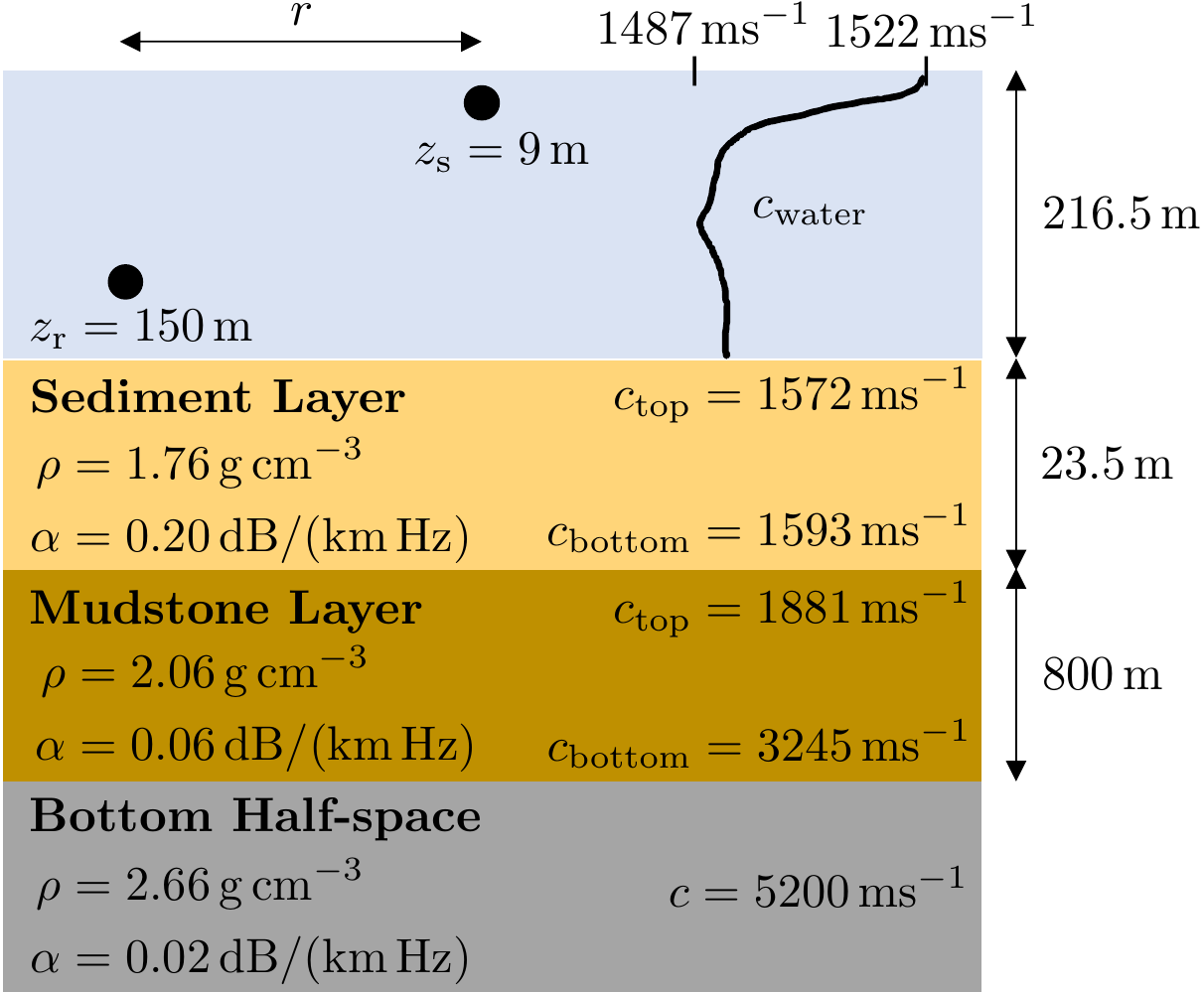}
    \caption{Acoustic environment used for performance evaluation of the proposed AUV navigation method. In the considered range-independent shallow-water scenario, the sound speed profile and seafloor properties are adopted from the SWellEx-96 experiment event S5.\vspace{-5.5mm}} 
     \label{fig:simEnv}
   \end{minipage}
 \end{figure}

The simulation environment follows the geoacoustic model from the S5 event of the SWellEx-96 experiment with the vertical line array (VLA)\cite{swellex:E96} (Fig.~\ref{fig:simEnv}). The source depth was chosen as $z_s \rmv = \rmv 9$. This is the same depth as the towed source at the shallow depth in the SWellEx-96 experiment above the thermocline. Rouseff and Spindel \cite{RouSpi:J02} discuss the impact of different factors, including the source depth, on the ``distribution'' of the WI. The receiver depth was arbitrarily chosen close to the 34\textsuperscript{th} element of the VLA at a depth of $z_r \rmv = \rmv 150$ m. The acoustic signals are simulated with an acoustic modal simulator, KRAKEN, which generates the pressure field in the range and frequency\vspace{-1mm} domain.
% Fig. Simulation Environment
\subsection{Range Rate and Range Computation}\label{sec:compDataDetails}
\vspace{-.5mm}
The range rate computation uses 4 min long segments, while the range computation uses a 10 min long signal. In the range rate computation, simulation and real data use the signals at $109$ Hz. For the range computation, the acoustic signals of four tones ($109, 127, 145, 163$ Hz) from a source at a shallow depth are used in both simulation and real data. When the first range rate measurement becomes available, another  $10$ min long acoustic recording is necessary to compute the range. Recall that using the $4$ min long spectrogram, the first range rate can be computed after $2$ min. Hence, the first range measurement becomes available after $12$ min.

\subsubsection{Simulation}\label{sec:simResult}
Monte Carlo experiments with different signal-to-noise ratios (SNRs) of the acoustic recording under the scenario described in Sec.~\ref{sec:simScenario} are performed to characterize the performance of the range rate computation and the range computation. The SNRs considered are $3$, $5$, $10$, $15$ and $20$ dB, and $300$ Monte Carlo experiments were conducted for each SNR. Here, circularly symmetric Gaussian noise is added to the source signal.

\begin{figure}[t]
   \centering
   \begin{minipage}{0.22\textwidth}
      \centering
      \centerline{\includegraphics[width=\linewidth]{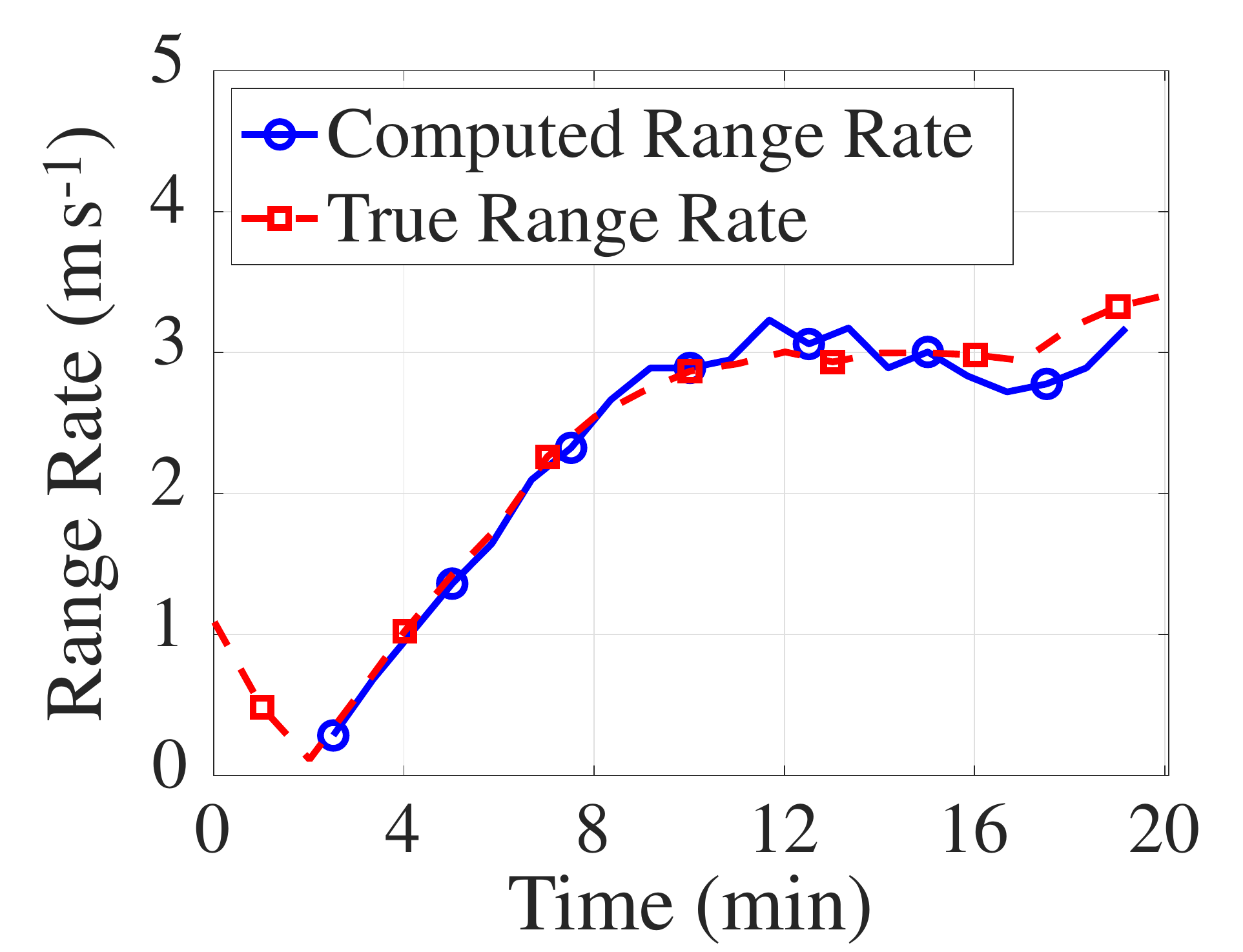}}
      \vspace{-2mm}
      \centerline{\scriptsize (a)}   
   \end{minipage}
   \begin{minipage}{0.22\textwidth}
      \centering
      \centerline{\includegraphics[width=\linewidth]{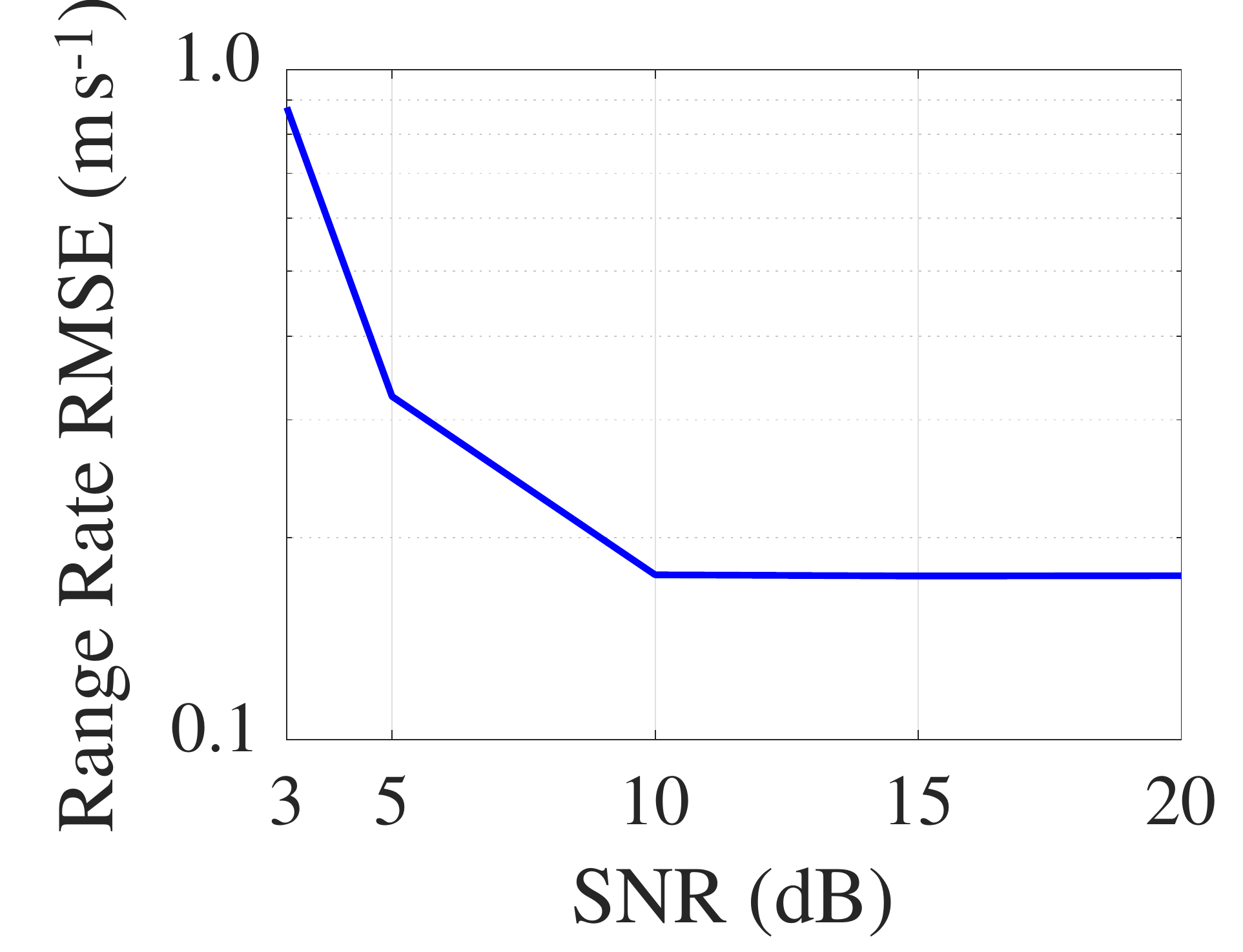}}
      \vspace{-2mm}
      \centerline{\scriptsize (b)}
   \end{minipage}
   \hfill
   \begin{minipage}{0.22\textwidth}
      \centering
      \centerline{\includegraphics[width=\linewidth]{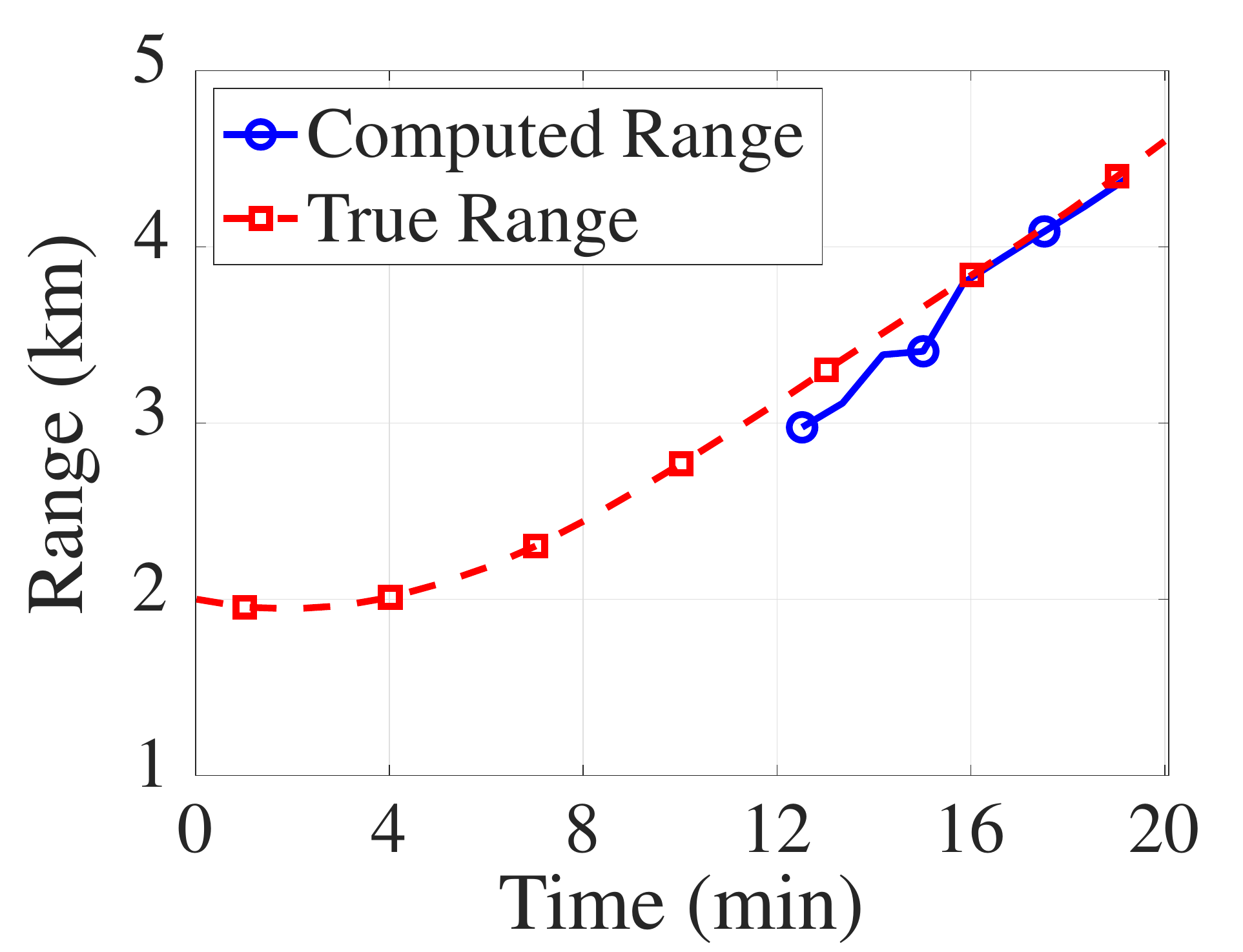}}
      \vspace{-2mm}
      \centerline{\scriptsize (c)}
   \end{minipage}
   \begin{minipage}{0.22\textwidth}
      \centering
      \centerline{\includegraphics[width=\linewidth]{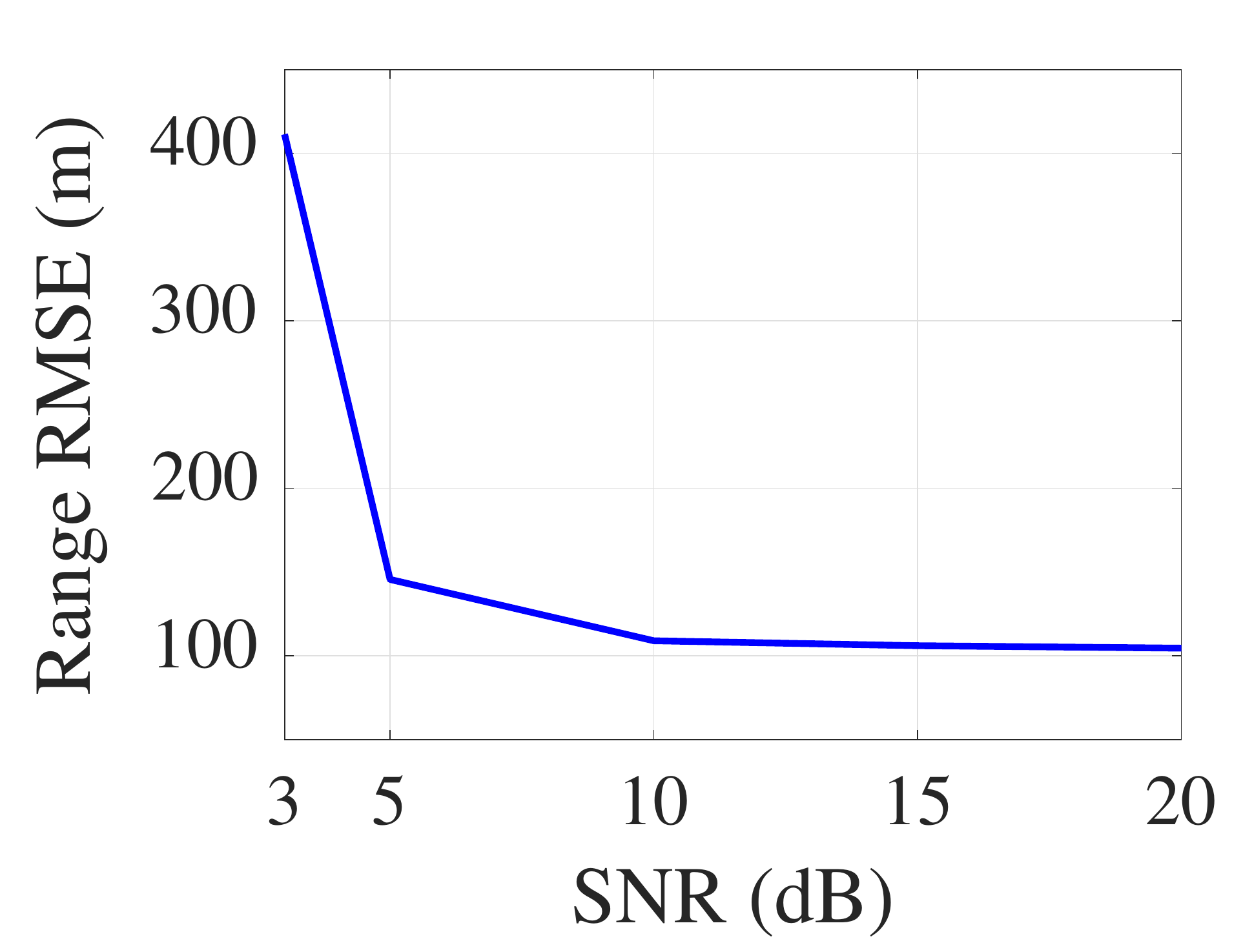}}
      \vspace{-2mm}
      \centerline{\scriptsize (d)}
   \end{minipage}
   \caption{An example of range rate computed in a simulated scenario is shown in (a). This range rate computation requires data with a duration of $4$ min. The range rate RMSE at higher SNRs is biased due to the limited resolution and the approximations made by the range rate computation method. The RMSE of the computed range rate as a function of the SNR is presented in (b). An example of a range computed in a simulated scenario is shown in (c). Since the range computation is feasible $10$ min after the range rate computation becomes available, the curve for the computed range starts after $12$ min. The RMSE of the computed range, which relies on the computed range rate as a function of the SNR, is depicted in (d).\vspace{-5.5mm}}
   \label{fig:rrrRMSE}
\end{figure}

The range rates computed from the simulated data are compared to the true range rate at every 30 s interval, and their root-mean-square errors (RMSEs) sampled for each SNR are calculated (Fig.~\ref{fig:rrrRMSE}(a) and (b)). The RMSE converges to a biased value of ${\sim}0.17$ m$\,$s\textsuperscript{-1} as the SNR increases. This bias is introduced mainly by two factors. The first factor is the limited time interval sampling, which results in a range rate resolution of $0.11$ $\mathrm{m\,s^{-1}}$. The second factor is a combination of the wavenumber variation \cite{RakKup:J12} and the mismatch between the sound speed of the water, $v$, and the horizontal phase speed, $\overline{v_p}$. 

Based on these computed range rates, the ranges between the source and receiver are calculated as described in Sec.~\ref{sec:rEst} (Fig.~\ref{fig:rrrRMSE}(c)). The computed WI for the simulated environment is $z_{\beta,\mathrm{sim}}\rmv = \rmv 1.11$. The computed ranges are also compared to the true range at every 30 s interval to calculate the range RMSE over the Monte Carlo experiments for each SNR (Fig.~\ref{fig:rrrRMSE}(d)).

% Fig. Tonal Signal Example
\begin{figure*}[t]
   \begin{minipage}{0.33\textwidth}
      \centering
     \centerline{\includegraphics[width=\linewidth]{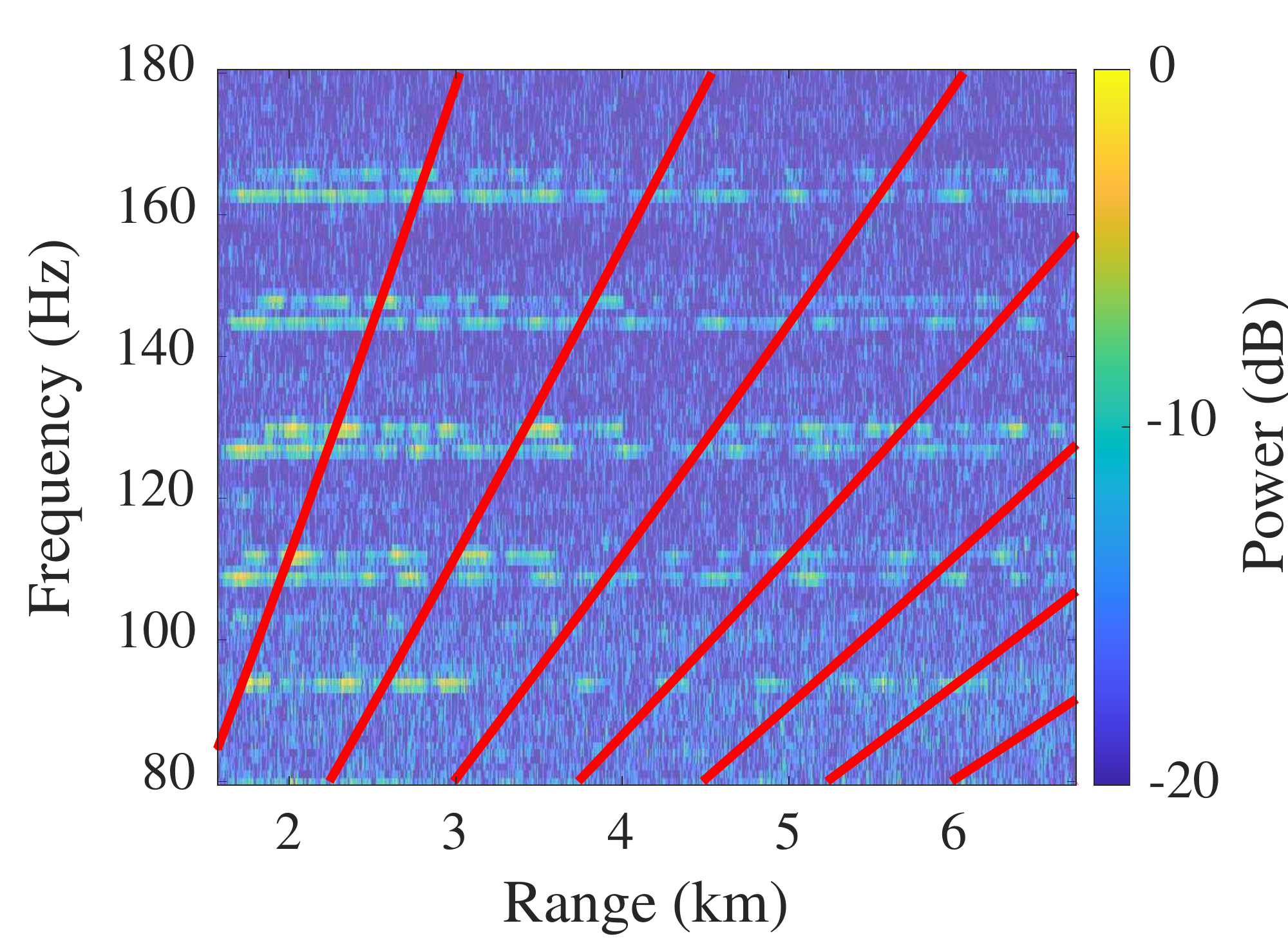}}
      \vspace{-2mm}
      \centerline{\scriptsize (a)}\medskip   
   \end{minipage}
   \begin{minipage}{0.33\textwidth}
      \centering
      \centerline{\includegraphics[width=\linewidth]{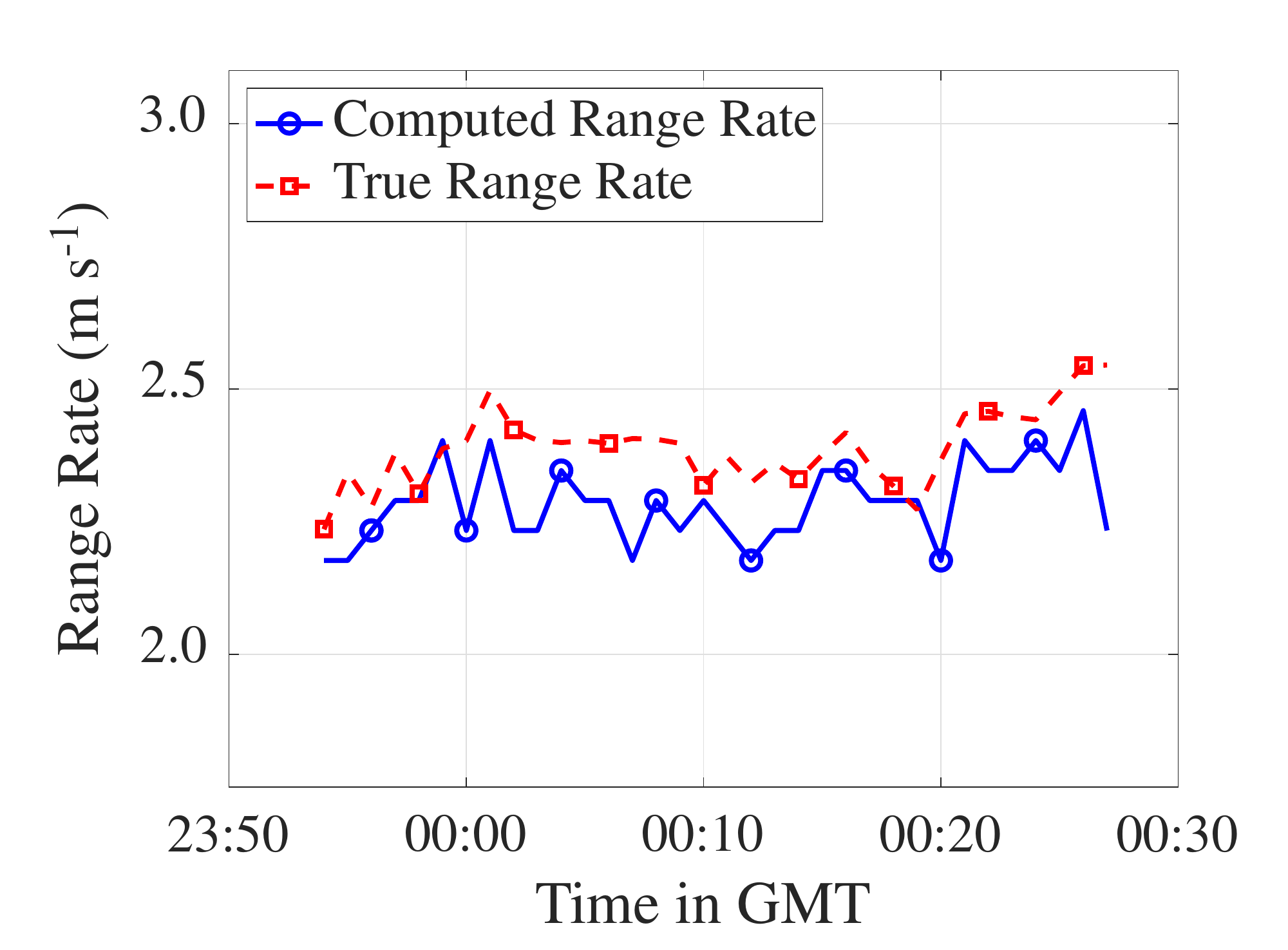}}
      \vspace{-2mm}
      \centerline{\scriptsize (b)}\medskip   
   \end{minipage}
   \vspace{-1mm}
   \begin{minipage}{0.33\textwidth}
      \centering
      \centerline{\includegraphics[width=\linewidth]{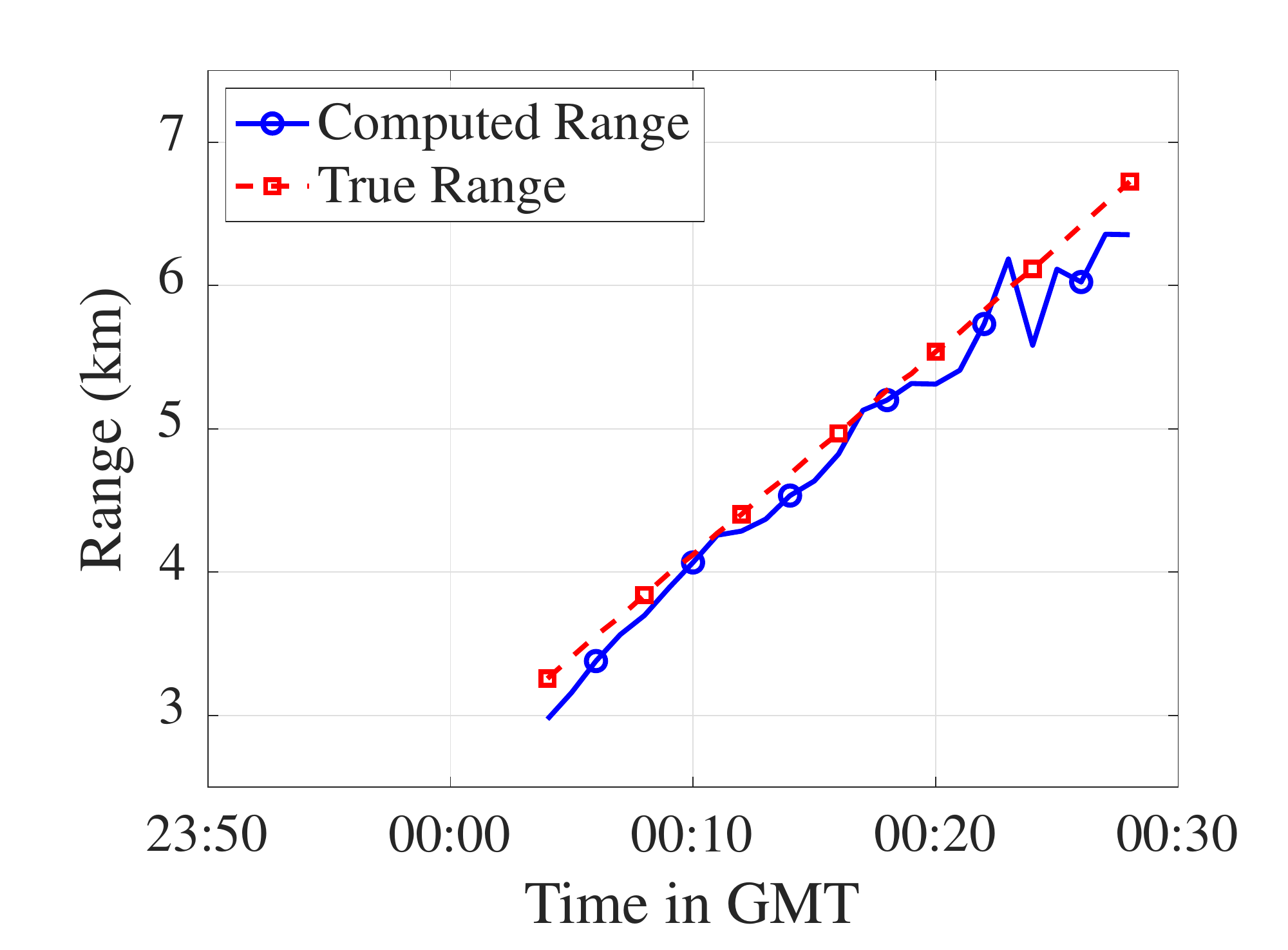}}
      \vspace{-2mm}
      \centerline{\scriptsize (c)}\medskip
   \end{minipage}
   \vspace{-3mm}
   \caption{A spectrogram with tonal signals from the moving source is shown in (a). This spectrogram relies on the acoustic data recorded by the first element of HLAS during the SWellEx-96 experiment event S5. The red lines indicate the striations related to $\beta=1.15$. A single tone at $109$ Hz is used for computing the range rate measurement, and four tones at frequencies $109, 127, 145$, and $163$ Hz are used to compute the range measurements. The range rate and range measurements are compared with their respective ground truth in (b) and (c). The average SNR of the acoustic signal is $8$ dB in the considered dataset.\vspace{-5mm}}
   \label{fig:swellexResult}
 \end{figure*}

\subsubsection{Real Data}
The geoacoustic properties are equivalent to the simulation environment, except for the seafloor and receiver depths. The receiver is located on the seafloor, which is 198 m deep. The noise variance for the SNR calculation is obtained as the power at which the tones were not present.

The computed range rates are in good accordance with the true ranges rate but appear to be biased by $0.10$ m$\,$s\textsuperscript{-1} (Fig.~\ref{fig:swellexResult}(b)). On top of the two factors for the bias discussed in Sec.~\ref{sec:simResult}, there is also a model mismatch. The bathymetry along the event S5 slowly changes; therefore, it is a mildly range-dependent environment. The true range rate was computed by taking a moving average of the range difference between the receiver and the ship's location, which was recorded with a GPS every minute. In addition, the computed range rate considers the phase offset discussed in \cite{WanWanDu:J17}. In particular, given the sampling frequency of $3276.8$ Hz, the phase offset due to taking a segment of length $3276$ samples is $0.36$ m$\,$s\textsuperscript{-1}, which is removed from the computed range rates.

Using $z_{\beta,\mathrm{real}}\rmv = \rmv 1.15$, which was computed using the same data, the range RMSE compared to the actual range is $217$ m (Fig.~\ref{fig:swellexResult}(b)). The average SNR of the data used is calculated to be $8$ dB. Note also that the range computed using the average of the moving averaged true range rate ($2.39$ m$\,$s\textsuperscript{-1}) yields an RMSE of $104$ m. In comparison, the range RMSE based on the simulation is approximately $109$ m at average SNR of $10$ dB in Fig.~\ref{fig:rrrRMSE}. Given the bias in the computed range rate, it is promising that the range computed using the SWellEx-96 event S5 data is close to the expected error. 

Even though no real acoustic recordings from an AUV are available in this work, the following tracking simulation results show that our proposed method has a strong potential for bounding the error for self-localization in underwater navigation scenarios with just a single hydrophone in addition to the sensors for a dead-reckoning\vspace{-1.5mm} navigation.

\subsection{Navigation}
\vspace{-.5mm}
$300$ Monte Carlo experiments of the given the 20 min long scenario were performed to characterize the position error and covariance. The average SNR of the acoustic signal is $12$ dB. Assuming that the AUV has traveled based on dead-reckoning for some time, the initial position estimate is sampled from a normal distribution with a standard deviation of $500$ m. The corresponding WI parameter is computed as $z_{\beta,\mathrm{sim}} \rmv = \rmv 1.11$. To avoid particle degeneracy\cite{LiBolDju:J15}, $500,000$ particles were used, and a Gaussian kernel with a standard deviation (STD) of $10$ m was added to the position states of the particles at every resampling stage for roughening. Using a MacBook Pro with an Apple M1 Pro chip and $32$ GB memory, the particle-based estimation at each time step takes a median value of 0.11 s.

Each measurement noise STD reflects the error specifications of real sensors used in the AUV navigation. The STD of the heading and the speed over the ground are modeled after OceanServer OS5000\cite{os5000:D10} and Teledyne RD Instruments Explorer Doppler Velocity Log\cite{explorer:D13}, respectively. 

\begin{figure*}[t]
   \centering
   \begin{minipage}{0.325\textwidth}
      \centering
      \centerline{\includegraphics[width=\linewidth]{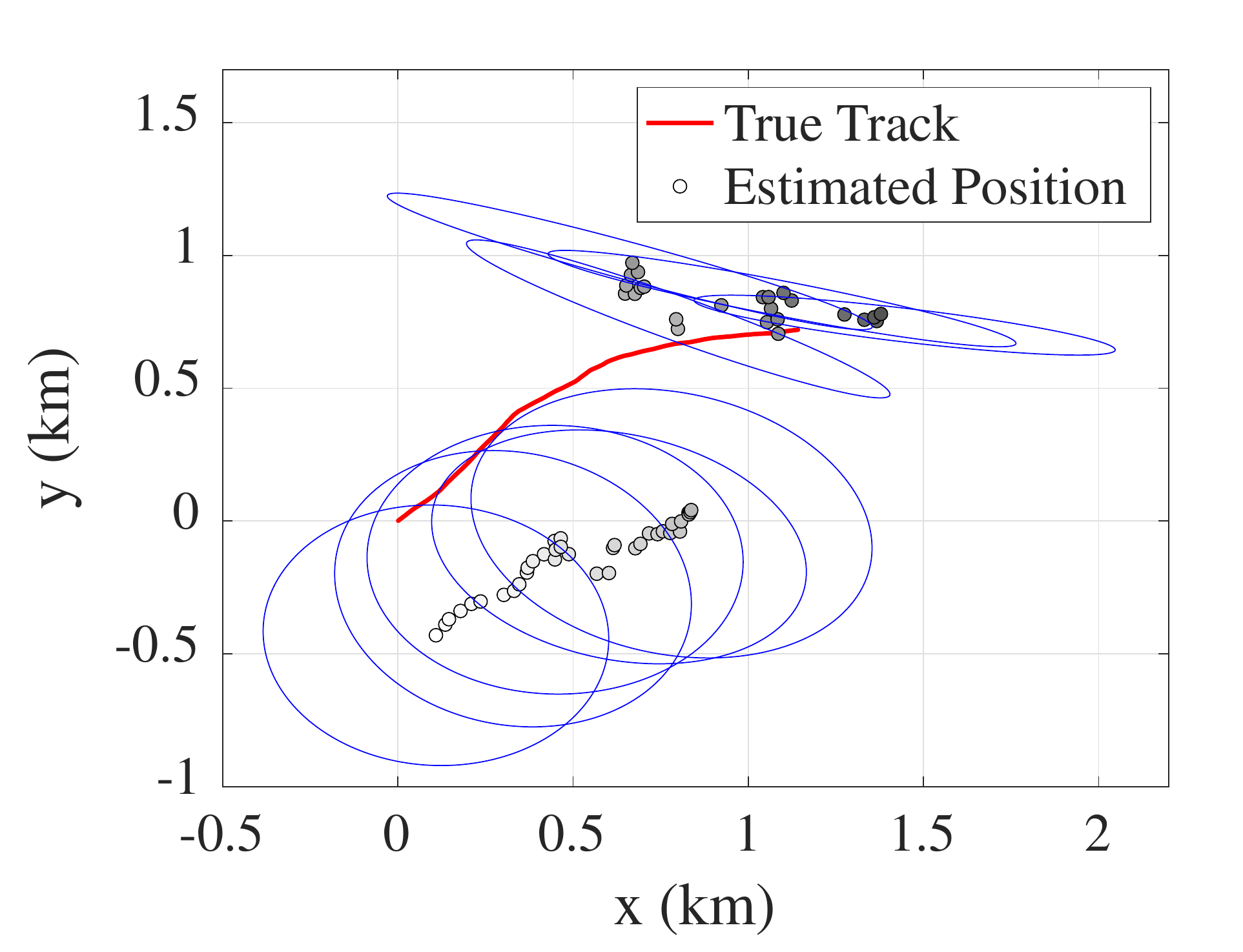}}
      \vspace{-2mm}
      \centerline{\scriptsize (a)}\medskip
    \end{minipage}
    \begin{minipage}{0.325\textwidth}
      \centering
      \centerline{\includegraphics[width=\linewidth]{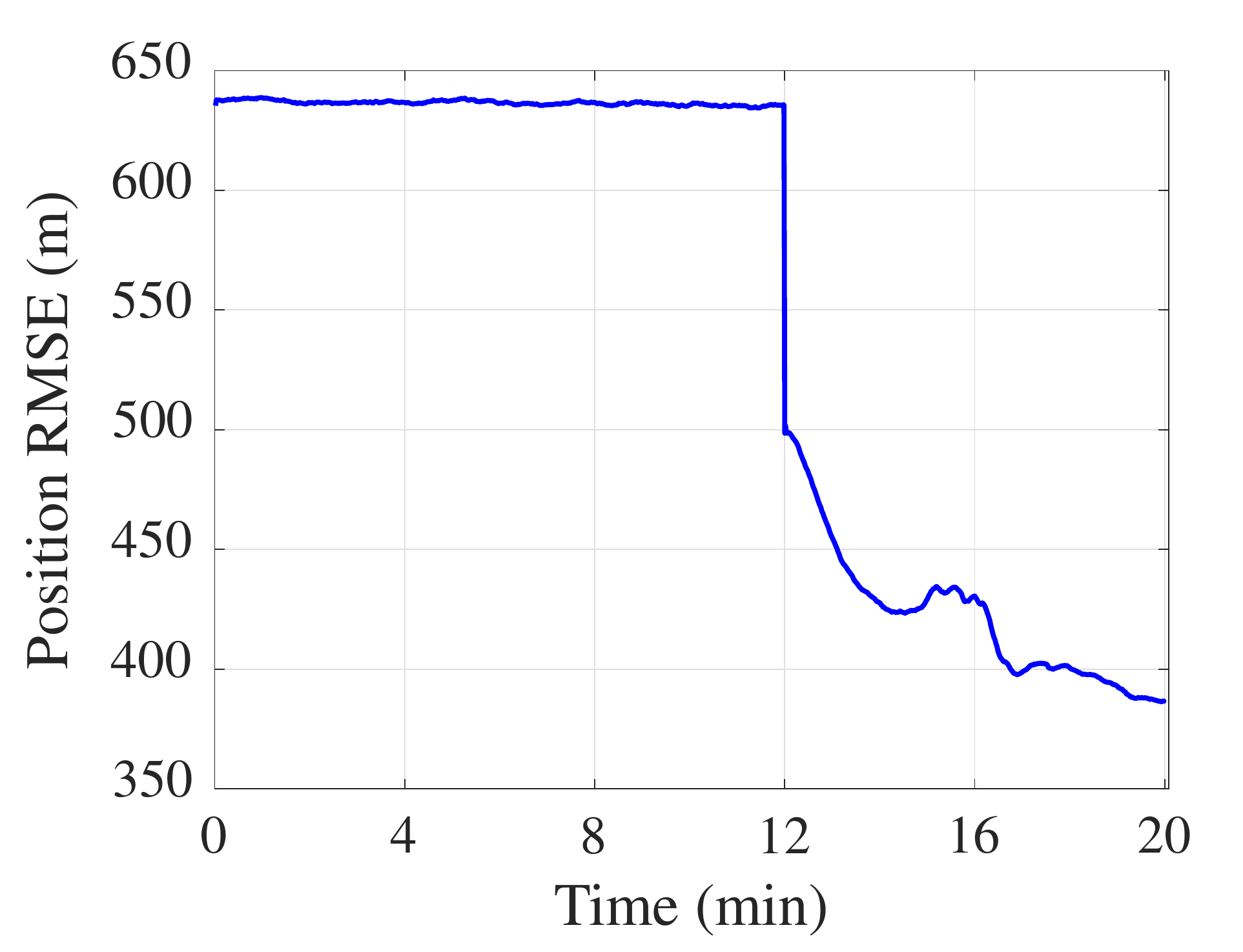}}
      \vspace{-2mm}
      \centerline{\scriptsize (b)}\medskip   
   \end{minipage}
   \begin{minipage}{0.325\textwidth}
      \centering
      \centerline{\includegraphics[width=\linewidth]{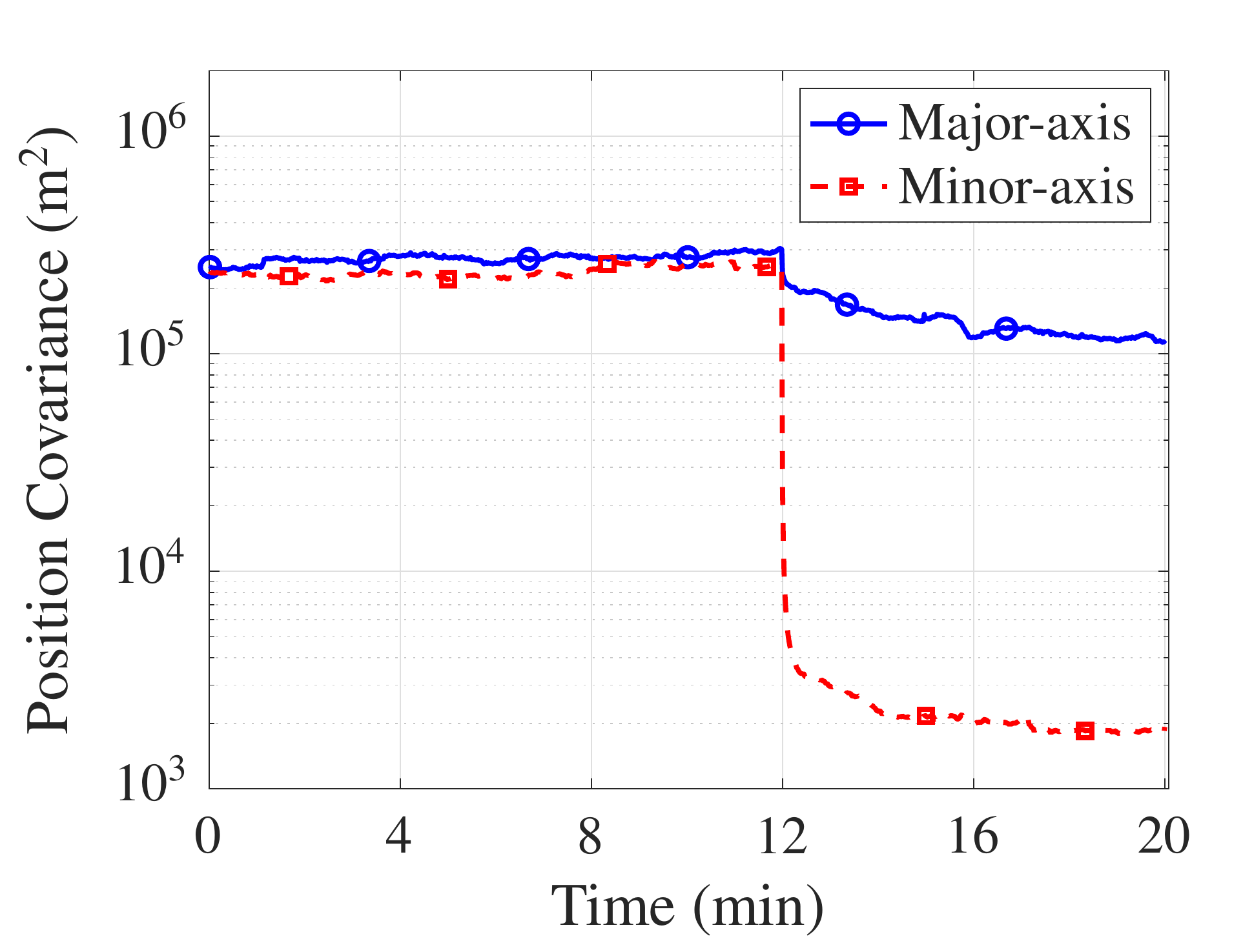}}
      \vspace{-2mm}
      \centerline{\scriptsize (c)}\medskip
   \end{minipage}
   \vspace{-3mm}
   \caption{An example of an AUV navigation scenario is shown in (a). This scenario is related to a single simulation run. Evolution in time is indicated by the shade of dots that show position estimates. The error ellipses represent the sample standard deviation of the position estimates along its major and minor axes. It can be seen that the ellipses become thin once the navigation filter incorporates the range measurements. The RMSE of the position estimates is compared to the true positions in (b). The average sample covariance of the particles representing the posterior distribution of the AUV position is presented in (c). Both (b) and (c) are based on $300$ Monte Carlo experiments. The strong reduction in RMSE and the covariance at $12$ min indicates the beginning of the WI-based ranging.\vspace{-5mm}}
   \label{fig:posCovRMSE}
\end{figure*}

As discussed above, the first range measurement can be computed after $12$ min of acoustic data are available. As a result, sharp drops in the position error and the sample covariance of the particles after $12$ min are observed in Fig.~\ref{fig:posCovRMSE}. Furthermore, the navigation simulation results show that the position error is reduced once incorporated and keeps it\vspace{-2.5mm} bounded.

% Fig. MC Results

% ----------------------------------------------
% ----------- Sec.~V: Conclusion ------------
% ----------------------------------------------
\section{Conclusion}
\vspace{0mm}
We developed a Bayesian method for the navigation of autonomous underwater vehicles (AUVs) in shallow water. Our approach relies on passively recorded signals from acoustic sources of opportunities (SOOs).  A measurement of the range with respect to the SOO is extracted from the spectrogram of a single hydrophone by exploiting striation patterns described by the waveguide invariant. A particle-based navigation filter processes the resulting range measurements and the AUV's internal velocity and heading measurements. As a result, the position error, which would otherwise increase over time, can be bounded. The ability to compute the range rate and range measurements from the pressure field measured using a single hydrophone is demonstrated on real data from the SWellEx-96 experiment.  The achieved ranging accuracy shows potential for real-world applications. In particular, by adding just a single hydrophone to a navigation system that only relies on dead-reckoning, the system can limit its position uncertainty if a suitable SOO is available. Future research directions include an extension to tracking of SOOs \cite{MeyKroWilLauHlaBraWin:J18,JanMeySny:J23,MeyWil:J21} and model adaptation using deep\vspace{0mm}  learning \cite{LiaMey:J23}.

\section*{Acknowledgement}
\vspace{-.5mm}
This work was supported by the Defense Advanced Research Projects Agency (DARPA) under Grant D22AP00151. We thank Dr. H.~Akins, Dr.~A.~H.~Young, Prof.~S.~T.~Rakotonarivo, Prof.~A.~H.~Harms and Prof.~W.~A.~Kuperman for interesting discussions and comments\vspace{-2mm}.

\renewcommand{\baselinestretch}{.983}
\selectfont
\bibliographystyle{IEEEtran}
\bibliography{IEEEabrv,StringDefinitions,SALPapers23,SALBooks23,Temp}

\end{document}

%% file: LatexInclusion/defmetric.tex
\newcommand{\bd}{\begin{description}}
\newcommand{\ed}{\end{description}}
\newcommand{\be}{\begin{enumerate}}
\newcommand{\ee}{\end{enumerate}}
\newcommand{\bi}{\begin{itemize}}
\newcommand{\ei}{\end{itemize}}
\newcommand{\bl}{\begin{list}}
\newcommand{\el}{\end{list}}
\newcommand{\bt}{\begin{tabbing}}
\newcommand{\et}{\end{tabbing}}

\definecolor{BLUE}{rgb}{0,0,1}